  \documentclass[12pt]{article}
\usepackage{amssymb,amsmath}
\usepackage{graphicx}
\usepackage{epsfig}
\usepackage{epstopdf}
\usepackage{latexsym}
\usepackage{psfrag}
\usepackage{booktabs}
\usepackage{bbm}

\usepackage{color}
\usepackage[
      colorlinks=false,
      linkcolor=darkblue,  
      urlcolor=blue,    
      filecolor=blue,     
      citecolor=red,
linktocpage=true,
      pdfstartview=FitV,
      bookmarksopen=true    
      ]{hyperref}

\ifpdf
\fi

\numberwithin{equation}{section}

\def\coeff#1#2{\relax{\textstyle {#1 \over #2}}\displaystyle}

\def\IR{\mathbb{R}}
\def\ZZ{\mathbb{Z}}

\def\cB{{\cal B}}

\def\cL{{\cal L}}
\def\cM{{\cal M}}
\def\cN{{\cal N}}
\def\cO{{\cal O}}

\def\cQ{{\cal Q}}

\def\cS{{\cal S}}

\def\Neql#1{{\cal N}\!=\!{#1}}

\def\ie{{\it i.e.}}

\def\bL{{\bf L}}
\def\bK{{\bf K}}\def\BR{{\Bbb R}}\def\BE{{\Bbb E}}
\def\BZ{{\Bbb Z}}
\def \p{\partial} \def \half{\frac{1}{2}}

\definecolor{cardinal}{rgb}{0.6,0,0}
\definecolor{darkgreen}{rgb}{0,0.5,0}
\definecolor{golden}{rgb}{0.92, 0.7, 0}
\definecolor{midnight}{rgb}{0, 0, 0.5}
\definecolor{darkblue}{rgb}{0.2, 0, 0.8}

\begin{document}  

\begin{titlepage}
\rightline{DAMTP-2013-11}
\rightline{IPHT-t13/059}
\bigskip
\bigskip

\begin{center} 
{\Large \bf  Global Structure of  Five-dimensional BPS
Fuzzballs }

\bigskip
\bigskip 

{\bf G.W. Gibbons${}^{(1)}$ and N.P. Warner${}^{(2,3,4)}$ \\ }
\bigskip
${}^{(1)}$
D.A.M.T.P., University of Cambridge \\ Wilberforce Road, Cambridge CB3 0WA, U.K.

\bigskip

$^{(2)}$ Department of Physics and Astronomy \\
University of Southern California \\
Los Angeles, CA 90089, USA\\

\bigskip
$^{(3)}$ Institut de Physique Th\'eorique, CEA Saclay \\
CNRS-URA 2306, 91191 Gif sur Yvette, France\\

\bigskip
$^{(4)}$ Institut des Hautes Etudes Scientifiques \\
Le Bois-Marie, 35 route de Chartres \\
Bures-sur-Yvette, 91440, France

%
\end{center}

\bigskip

\begin{abstract}

\noindent  We describe and study families of BPS microstate geometries, namely, smooth, horizonless asymptotically-flat solutions to supergravity.  We examine these solutions from the perspective of  earlier attempts to find solitonic solutions in gravity and show how  the  microstate geometries circumvent  the earlier ``No-Go'' theorems.  In particular, we re-analyse the Smarr formula and show how it must be modified in the presence of non-trivial second homology.  This, combined with the supergravity Chern-Simons terms, allows the existence of rich classes of BPS, globally hyperbolic, asymptotically flat, microstate geometries whose spatial topology is the connected sum of $N$ copies of $S^2 \times S^2$ with a ``point at infinity'' removed.   These solutions also exhibit ``evanescent ergo-regions,'' that is, the non-space-like Killing vector guaranteed by supersymmetry is  time-like everywhere except on time-like  hypersurfaces (ergo-surfaces) where the Killing vector becomes null.
As a by-product of our work, we are able to resolve the puzzle of why some regular soliton solutions  violate the BPS bound:
their  spactimes do not admit a spin structure.

\end{abstract}

\end{titlepage}

  
\tableofcontents

\newpage
\section{Introduction:  Six Impossible things before breakfast }
\vskip -0.3cm
{\small \it Alice laughed.``There's no use trying,'' she said: ``one can't believe impossible things.''}
\hfill \break
{\small \it ``I daresay you haven't had much practice,'' said the Queen. ``When I was your age, I always did it for half-an-hour a day. Why, sometimes I've believed as many as six impossible things before breakfast''} 

\rightline{\small {\it    Through the Looking-Glass}, Lewis Carroll}
\smallskip

Much work over the last eight years  has been  devoted to constructing candidate `microstate geometries,' sometimes referred to as `supergravity fuzzballs,' with a view to resolving  some long-standing puzzles in the quantum theory of black holes.  The original idea of fuzzballs as a possible semi-classical geometric description of black-hole microstates originated from the work of Mathur and collaborators on two-charge systems.  (For  reviews of work on the two-charge systems, see \cite{Mathur:2005zp,Mathur:2005ai,Mathur:2008nj}.)  In the F1-P duality frame,  the states of the two-charge BPS system can be described by momentum modes on a fundamental string and so can easily be analysed and counted within the underlying conformal field theory.  In supergravity, these solutions can be given a semi-classical description in which the string source with momentum waves is represented by an arbitrary shape profile for the string source.  In addition to the F1-P electric charges, there is also a dipolar F1 charge and angular momentum along the string profile.  If one transforms this to the duality frame in which it has D1-D5 charges then the dipole charge becomes that of  a Kaluza-Klein monopole (KKM).  The corresponding supergravity solution encodes that same semi-classical data as its F1-P counterpart but is, in fact, completely smooth in six dimensions \cite{Lunin:2001fv,Lunin:2002iz}.  These are the original, prototypical microstate geometries: they are smooth, horizonless, asymptotically flat solutions with the same boundary conditions at infinity as a two-charge black hole.  Mathur and collaborators went on to demonstrate that these fuzzball geometries could reproduce many of the properties of BPS black holes and of near-BPS black holes (see \cite{Lunin:2001jy, Lunin:2002iz, Mathur:2005zp} for further discussion of this).

The problem with the two-charge system is that the corresponding black holes do not have a classical horizon:  they only have a ``stretched'' or ``effective horizon''   at the  Planck or string scale.  This means that the smooth, classical configurations that are supposed to account for the black-hole entropy are at, or very near, the limit of the supergravity approximation.  In spite of this, Sen has made some very interesting links between the microstate counting and the Bekenstein-Hawking entropy for the two-charge system  \cite{Sen:1995in}.  This work was thus extremely  suggestive but it really required the later analysis of  Strominger and Vafa \cite{Strominger:1996sh} to make a more compelling connection, at least at vanishing string coupling.  The important difference was that Strominger and Vafa did the analysis for the three-charge black hole in five dimensions whose horizon area is much larger than the Planck scale.  Similarly, the work of Mathur and collaborators on two-charge microstate geometries is remarkably detailed, very interesting and suggestive but it really requires the development of the three-charge microstate geometries in order  to have confidence in the validity of the supergravity approximation.

The first smooth, three-charge BPS microstate geometries in which all three charges are large were constructed in \cite{Giusto:2004kj,Bena:2005va, Berglund:2005vb}.  These solutions replaced the singular electric sources by smooth topological fluxes.  They could also be given a rather attractive interpretation in terms of  brane sources that have undergone a geometric transition.  The systematic study of these solutions, their generalizations and the classes of non-BPS counterparts been a very active and fruitful area of research in the last few years.  

Perhaps the simplest of these microstate geometries are  asymptotically flat solutions of the equations of five-dimensional supergravity theories.  However, there are  richer classes of candidate microstate solutions that suffer from singularities when considered as five-dimensional spacetimes but  yield  smooth, non-singular,  but no longer  asymptotically flat, solutions of supergravity theories when lifted six space-time dimensions.  

The five-dimensional microstate geometries represent a remarkable class of  smooth, geodesically complete, asymptotically flat, stable spacetimes without horizons \cite{Giusto:2004kj,Bena:2005va,Berglund:2005vb,Saxena:2005uk}. (For  reviews of work on the three-charge systems, see \cite{Bena:2007kg,Skenderis:2008qn, Balasubramanian:2008da, Chowdhury:2010ct}.)  Their existence is rather surprising for  a number of reasons.   First, one might think of them as the gravitational analogue of solitons in flat-space field theories, and no such solutions are known in four-dimensional supergravity. The best  approximation to the soliton concept is the set of black-hole solutions  that are non-singular outside an `extreme'  or `degenerate'  event horizon but which harbour singularities inside.  Going back to the earliest days of General Relativity, there are a number of \lq No-Go \rq  results  \cite{Serini,Einstein,EinsteinPauli,Breitenlohner:1987dg,Carter2}, which exclude completely non-singular soliton solutions that are  regular in  four spacetime dimensions.  

For stationary metrics, one can establish no-go theorems\footnote{This is the first {\it impossible thing}.} using the Smarr formula,  which relates the Komar mass to densities that are to be integrated over interior boundaries.  We will review these ideas in Section \ref{Smarr1}. These boundaries are usually either singularities or horizons and so, if one has a smooth, horizonless, asymptotically-flat  solution one might naively expect that the mass must  be zero and hence the solution is necessarily trivial.  As we will discuss in some detail, five-dimensional supergravity is able to evade this conclusion because it has Chern-Simons interactions and the spatial sections may have non-trivial second homology.   As a result,  the one-form potentials may not be globally defined leading to extra bulk terms in the Smarr formula. 

Thus microstate geometries require non-trivial homology in the space-time.  BPS, or supersymmetric, geometries are necessarily stationary, with a non-space-like Killing vector, which however can be null on hypersurfaces. Supersymmetry also requires that the metric, $h_{\mu \nu}$, restricted to directions orthogonal to the Killing
vector must be conformal to a hyper-K\"ahler metric, $\hat h_{\mu \nu}$. Hyper-K\"ahler metrics are necessarily Ricci flat.   This presents a further challenge because a theorem of Schoen, Yau
 \cite{Schon:1979rg} and Witten \cite{Witten:1981mf} states that the only
 complete, non-singular, Ricci-flat, Riemannian manifold that is
 asymptotically Euclidean is, in fact, Euclidean space.  In particular, there is no topology\footnote{This is the second {\it impossible thing}.}. Again, five-dimensional
 supergravity dodges this bullet and does so because the condition
 that the complete space-time must be a smooth, Lorentzian
 five-manifold is weaker than requiring that the spatial base, $\cal B$, equipped with either
$h_{\mu \nu}$ or $\hat h_{\mu \nu}$ also be a complete, non-singular,
asymtotically-Euclidean,  Riemannian manifold. The base $\cal B$ does in fact admit  a  complete, non-singular
asymtotically-Euclidean,  Riemannian metric but it is neither $h_{\mu \nu}$ nor $\hat h_{\mu \nu}$ and it is not Ricci flat.

Since regularity of the five-dimensional, Lorentzian manifold 
does not require that the hyper-K\"ahler metric, $\hat h_{\mu \nu}$ to
be positive-definite, one may   allow it to be 
apparently pathological:  It can be `ambi-polar,' that is, the
signature can change from $+4$ to $-4$ with apparently  singular
intervening surfaces.  The miracle is that the `warp factors' of the time fibration and the angular momentum vector can convert this into a smooth, Lorentzian five-manifold.  The hypersurfaces on which the signature of   $\hat h_{\mu \nu}$  changes sign correspond to places where the Killing field becomes null.  Away from these hypersurfaces  the conformally related metric, $h_{\mu \nu}$, is positive definite but becomes singular when the Killing vector becomes null.  In contrast to the  usual situation in which this surface is a null hypersurface and thus a Killing horizon, in the present situation it is a timelike hypersurface.  In fact, this surface is a novel, and hitherto unencountered, phenomenon of an ergo-surface with no bulk ergo-region: an evanescent ergo-region.

From the perspective of string theory, microstate geometries also seemed an impossibility\footnote{This is the third {\it impossible thing}.} within the validity of any supergravity approximation.  Fifteen years ago, Horowitz and Polchinski \cite{Horowitz:1996nw,Damour:1999aw} observed that if one started from vanishing string coupling then stringy states, like all normal matter, will become more and more compressed as the string, and hence gravitational, coupling increases.  On the other hand, the size of a black-hole event horizon increases with the gravitational coupling and thus one should expect that the perturbative states that were counted in \cite{Strominger:1996sh} and seem to account for the entropy of a BPS black hole must necessarily be inside a horizon at any finite value of the string coupling.  Again supergravity  evades this conclusion because D-branes are solitonic objects whose tension is proportional to $g_s^{-1}$ and so they become floppier and floppier as the string coupling increases.  In particular, putting momentum and angular momentum modes on the right combination of D-branes can result in a supergravity configuration that grows at exactly the same rate as the would-be horizon size grows with $g_s$.

In five dimensions, a BPS black hole or black ring can only have a
macroscopic horizon if it has `three charges' and the horizon area
then grows as $\sqrt{Q_1 Q_2 Q_3}$.  While one can perform the
perturbative state counting for many D-brane systems, it was the
three-charge problem that was of crucial importance precisely because
it has a macroscopic horizon and thus the microscopic entropy of the
perturbative system  can be compared to the area of the horizon within
a valid supergravity approximation.   It is  also precisely for such
black holes that the  Chern Simons terms become non-trivial and the
microstate geometries scale with $g_s$ in the same way as the horizon.
However, activating the Chern-Simons form leads to another, possibly
daunting, practical issue:  This term has the form $F\wedge F \wedge
A$, rendering the generalized  Maxwell equations  non-linear.  
Thus microstate geometries might  have existed but  analytic solutions could  have proved impossible to find\footnote{This is the fourth {\it impossible thing}.}.  Even if some solutions could have been  found. or an existence theorem  proven, the non-linearities could have made the phase space structure impossibly difficult to analyze. 

However, yet another miracle happens:  The supergravity equations remain linear provided that they are solved  in the correct order \cite{Bena:2004de}.   One has to solve a sequence of linear equations, essentially involving only the Laplacian on the four-dimensional spatial base, and the non-linearities of the Chern-Simons terms only generate quadratic source terms assembled from the solutions of earlier equations in the linear system. Thus, once one has found a suitable ambi-polar, hyper-K\"ahler base everything is linear and families of solutions can be assembled by  superposition.

The only delicate part of the whole procedure is to make sure the ultimate solution is causally well-behaved.  In particular, this involves making sure that there are no closed time-like curves.  To do this one constructs complete four-dimensional hypersurfaces of constant time on which the induced metric is positive definite and everywhere non-singular.  This provides the solutions with a global Cauchy surface and precludes the existence of closed time-like curves. Technically, this involves imposing algebraic relationships between the cohomological fluxes and the moduli of the hyper-K\"ahler metric.  Intuitively, these algebraic relationships, or `bubble equations,' express the fact that there is a balance between the gravitational force that tends to contract topological cycles and the fluxes that tend to expand the cycles.  More generally,  there are many examples of smooth geometries that can be made in this way, but a `randomly assembled'  solution is most likely to be pathological.  As yet there are  no  general theorems about when smooth, causally well-behaved solutions can be constructed but some  heuristic suggestions were made in \cite{Bena:2007kg} and a much more specific proposal, called the ``split attractor conjecture,''  was proposed and investigated in  \cite{Denef:2000nb,Wang:2010sha}.

The first three-charge microstate geometries to be constructed with all three charges large were still extremely specialized:  they carried the same quantum numbers as a black-hole of zero horizon-area  \cite{Bena:2006is}.  That is, they typically had maximal angular momentum and the microstates described by these geometries appeared to be only marginally bound.  It became an open question as to whether any of these solitonic solutions could be arranged to look, asymptotically, like a true BPS black hole with what appears, from a distance, to be a macroscopic throat.  The solution to this problem came \cite{Bena:2006kb, Denef:2007vg, Bena:2007qc} from the class of microstate geometries called ``scaling solutions.''   From the perspective of the spatial base geometry, this appears to be a singular limit in which a cluster of homology cycles are blown down while the fluxes on them remain finite.  However, from the perspective of the five-dimensional geometry this limit corresponds to the opening of an $AdS$ throat and the smaller the cycles in the base geometry, the deeper the throat.  In this scaling geometry, the homology cycles limit to a finite size and the cross-section of the throat becomes a sphere whose radius  is  determined by the fluxes on the cycles that are scaling.  Thus the full geometry looks exactly like the $AdS$ throat of a typical extremal black hole or black ring except that it  `caps off' smoothly at some depth, determined by a scaling parameter, and the cap consists of finite-sized homology cycles carrying fluxes.  In this way, the microstate geometries have a scale that is set by the horizon size of the black hole.  
 
If a five-dimensional microstate geometry has a  $U(1)$ symmetry then it can be reduced to a four dimensions  and it generically becomes a four-dimensional, multi-centered black-hole solution.  The black-hole centers are located at the fixed points of the $U(1)$ action and the corresponding  sources may be thought of as fluxed D6-branes  \cite{Balasubramanian:2006gi}.  These solutions  play a very important role in the counting of black-hole entropy at weak coupling using quiver quantum mechanics \cite{Denef:2007vg}. Indeed, it was in this context that scaling solutions were first discovered and investigated \cite{Denef:2000nb, Denef:2002ru,Bates:2003vx} but it was not until later that the significance of scaling solutions for five-dimensional microstate geometries was discovered \cite{Bena:2006kb, Denef:2007vg, Bena:2007qc}. 

Classically, the depth of the $AdS$ throat in a five-dimensional scaling solution is set by a free parameter but semi-classical quantization of the moduli space of solutions limits the depth of the throat in a manner that depends upon the charges.  This has the interesting consequence that quantum effects are wiping out large scale naively-classical geometries\footnote{This is the fifth {\it impossible thing}.}:  the scales of much deeper throats are classically large,  certainly much larger than Planck scale,  and the supergravity approximation remains valid and yet quantum effects are removing such macroscopic structures \cite{Bena:2007qc,deBoer:2008zn,deBoer:2009un}.

Another  of the apparently impossible things that emerged from these families of microstate geometries is the extent to which they might be ``typical.''  Obviously, from the perspective of quantum mechanics, any large classical configuration is a very coherent state but a more important issue is whether it is a state from a highly atypical sector of the quantum theory.  The fact that these solutions have a very long AdS throat means that this issue can be addressed directly by using the AdS/CFT correspondence \cite{Kanitscheider:2007wq,Skenderis:2007yb,Skenderis:2008qn}. Indeed each smooth bubbled geometry with a long AdS throat must be dual to a state in the black-hole CFT that underpinned the original state counting in \cite{Strominger:1996sh}.  One can compute the lowest energy fluctuations of a scaling geometry by looking at the longest wavelength fluctuations that can be accommodated within the throat and then use the red-shifts generated by the depth that is set by the semi-classical quantization of the geometry.  One finds \cite{Bena:2006kb, Bena:2007qc,deBoer:2008zn} that these lowest energy fluctuations have the same energy as the fundamental fluctuations of the dual CFT that was used to count the states of the black hole \cite{Strominger:1996sh}.   Thus the microstate geometries have deep $AdS$ throats and are dual to states in the ``typical sector'' of the conformal field theory\footnote{This is the sixth {\it impossible thing}.} and do not seem to represent unusual outlier states in the overall ensemble.

In conclusion, despite many potential obstructions, there are indeed viable BPS microstate geometries that are, by definition, smooth, horizonless solutions lying within the validity of the supergravity approximation to string theory and that look exactly like BPS black holes and black rings until one gets arbitrarily close to the horizon.  The purpose of the present paper is to give a brief introduction to these remarkable solutions and then provide a global analysis, focussing on their topology and how the construction evades the various geometrical threats to their existence.  

The plan of the paper is as follows.  In Section 2 we provide some historical context for our work by reviewing some of the earlier attempts to find  solitons in field theory.  In Section 3  we discuss issues of global structure and causality in five dimensions and we discuss the requirements on the metric that guarantee that the space-time is non-singular and globally hyperbolic and we give a simple example of an evanescent ergo-region.  In Section 4, we introduce the $\cN=2$ supergravity theory, coupled to two vector multiplets, that will be the basis of our analysis and we describe the metric Ansatz and BPS equations that are appropriate to the study of geometries that preserve the same supersymmetries as a BPS black hole.   Section 5 contains a careful analysis of the Smarr formula for $\cN=2$ supergravity in five dimensions and, in particular, we  examine precisely how topological contributions can arise in the Smarr formula and thus how microstate geometries can evade the ``no-go'' theorems.   In  Section 6 we review families of microstate geometries and analyze two examples in detail. Another non-BPS example is discussed in Section 7.   Using the methods developed earlier,  we are able to resolve  the puzzle why some  regular soliton solutions  violate the BPS bound. It turns out that these particular non-BPS space-times do not admit a spin structure.   The topology of the BPS solutions solutions is discussed in Section 8 and our conclusions may be found in Section  9.

\section{A brief history of solitons versus particles}

In our introduction we reviewed the history of microstate geometries within string theory.  In this section we shall place this within the general context of efforts to replace point particles by smooth, essentially geometrical configurations, now often called solitons. In this way we hope to set the results described in the introduction and the results of the present paper in the context of a great deal of activity going back to at least the beginning of the last century:  the aim being to replace Boscovich's eighteenth century vision of point particles acted upon by the fields they generate \cite{Boscovich}. 

This programme became more urgent with the rise of classical electron theory \cite{Lorentz}  and  the realisation  that the self-energies of electrons, modelled as classical point particles, contributes to their inertia, but  is divergent.  Even if this problem could be circumvented, there remained the  
objection that in such theories of particles and fields, the equations of motion for the particles and for the fields need to be postulated separately.

Gustav Mie \cite{Mie}, in an attempt to to solve both problems turned to non-linear theories of electrodynamics, in which the classical self-energy 
$$
\int \half  {\bf D} \cdot {\bf E}\, d^3\, x
$$
of charged  particles could be be finite because  the electric displacement ${\bf D}$ is 
a non-linear function of the electric field ${\bf E}$. Another consequence of the non-linearity was the possibility
that the equations of motion of the finite-energy point singularities might follow from the non-linear field equations
without the need to be postulated separately, thus arriving at what was then referred to as a   ``Unitary Theory.''

With the advent of quantum mechanics, these attempts were, for a while, 
abandoned but the development  of Quantum Electrodynamics and 
the discovery that   vacuum fluctuations led 
to both divergent self-energy
and non-linear corrections to Maxwell's equations 
motivated Born to return to them, and develop
the most attractive version of the ideas now known as
Born-Infeld theory \cite{Born}. Born's programme continues to attract interest
in its own right  not least because of its relation
to strings ending on p-branes \cite{Fradkin:1985qd,Abouelsaood:1986gd,Strominger:1995ac,Townsend:1995af,
Callan:1997kz,Gibbons1}. It appears that the question  whether the equations   motion of its singular but finite energy classical point particle solutions
(BIons\footnote{The term {\it BIon} \cite{Gibbons1}  should be carefully  distinguished  from  {\it Soliton} which we take to be an {\it everywhere   non- singular}
solution of the classical equations of motion with finite total energy.
The standard example of a soliton in relativistic field theory 
is the `t Hooft- Polyakov monopole \cite{'tHooft:1974qc,Polyakov:1974ek}.}) follow rigourously from the equations of motion is still open  \cite{Kiessling1,Kiessling2,Kiessling3,Kiessling4,Kiessling5,Kiessling6,Speck,Chernitskii,Chruscinski}.

Einstein's construction of General Relativity, a classical non-linear theory par-excellence, was very early on seen to raise  the same questions for  gravitating particles, and gravitating bodies more generally. Two questions arose immediately.

YYY
  
\begin{itemize}

\item (1) Does the vacuum theory ({\it i.e.}   $R_{\mu \nu}=0$)     admit  everywhere non-singular finite energy ({\it i.e.}   asymptotically flat) solutions?

\item (2) 
 Do the equations of motion of gravitating bodies follow from the Einstein equations
($R_{\mu \nu} - \half  g_{\mu \nu} R= 8 \pi T_{\mu \nu}$)  without needing to be postulated separately?

\end{itemize}

Question 1 was answered initially in   the negative by Serini \cite{Serini} and 
with increasing precision by Einstein and Pauli \cite{Einstein,EinsteinPauli} and Lichnerowicz \cite{Lichnerowicz}.

Question 2  
was answered in the positive by Einstein,  Infeld and Hoffman \cite{EIH} \footnote{The basic point here is that the Bianchi identities imply  
(covariant) energy momentum conservation
$T^{\mu \nu}\,_{;\nu}=0$, which are the matter equations of motion.
For electro-dynamics, however, the analogue of the
Bianchi identities give a weaker result, current conservation
 $J^\mu\,_{;\mu} =0$.}.
 
 \bigskip

The development of gauge theory and the discovery of non-abelian  monopoles
in Yang-Mills-Higgs theory \cite{'tHooft:1974qc,Polyakov:1974ek} sparked off renewed interest
in classical models of particles or, in Coleman's happy phrase, classical lumps
and their quantum  descendants \cite{Coleman}.  It was then natural to
ask what were the analogues of these classical lumps in general relativity
\cite{Hajicek:1980mb,Gibbons82}. Of course it is rather clear that
given a stable finite energy soliton solution of a set of non-linear  field equations
in flat spacetime, such as a 't Hooft-Polyakov  monopole,
it should, when coupled to Einstein gravity, and 
if of a size large compared with its
Schwarzschild radius, remain stable. However as
its self-gravity becomes stronger its should undergo collapse to form
a black hole \cite{Gibbons:1990um,Ortiz:1991eu}.
Among such possibilities are self-gravitating   Q-balls
 \cite{Coleman:1985ki,Friedberg:1986tq}. The scalar fields supporting
Q-balls are   necessarily time dependent, however the metric
is time independent since it couples to the 
energy momentum tensor which is time independent for Q-matter.
 They are thus an interesting example of a
solution of Einstein's equations for which
the matter fields do not share all of the symmetries of the
metric. Examples of this phenonomeon occur in Einstein-Maxwell theory
and involve time or space-dependent duality transformations \cite{Tariq}.
In principle could occur in supergravity theories.
However there is no  example known to us which is asymptiotically
flat. In the present paper we shall always assume that {\sl both} the matter
and the metric are time independent.

The examples discussed in the previous paragraph,
while of possible importance in nature, cannot be thought of as owing
their  origin to  the properties of  gravity, since they can be found
in its absence. For the same reason, any topological characteristics
relate to the topology of the space of field configurations, not
to the topology of spacetime. Therefore, based on Lichnerowicz's theorem and its generalizations \cite{Carter1,Carter2} and Hawking's discovery of Black Hole Evaporation \cite{Hawking} 
it became clear that the only feasible candidate gravitational 
solitons in four dimensions were extreme black holes
\cite{Hajicek:1980mb,Gibbons82}  and this identification became
even more attractive if they were embedded in supergravity 
theories since extreme black holes in Einstein-Maxwell theory
are BPS. That is, they admit Killing spinors and fit into
supermultiplets \cite{Gibbons1981} 
and satisfy a Bogomolnyi bound \cite{Gibbons1982,Gibbons1982b}. 
The absence of true, singularity free  solitons 
in supergravity theories and an wide variety of related theories
was confirmed in \cite{Breitenlohner:1987dg} using a generalization
of the Smarr's formula  for the total mass of a spacetime
in terms of its charges  and angular momentum
and the area of any event horizon. In the absence of 
an horizon, Smarr's formula implies that the total energy $M$ vanishes
and hence by the positive energy theorem \cite{Schon:1979uj,Witten:1981mf}, the solution is trivial.       

Passing to five spacetime dimensions
and imposing the vacuum equations,  it is not difficult to see,  using the positive action
theorem  \cite{Schon:1979rg,Witten:1981mf} 
that Lichnerowicz's theorem remains true, provided one 
insists that the solution be asymptotically Euclidean
\cite{Gibbons1985} but if one relaxes this assumption
one obtains the Kaluza-Klein monopole \cite{Sorkin,Gross}
whose supersymmetric credentials were established in \cite{Perry}.
It should be noted however, that, even in the complete absence of matter, 
 supersymmetry will not prevent
gravitational collapse and black hole formation, in either five 
\cite{Bizon:2005cp}, or nine \cite{Bizon:2005af} dimensional flat spacetime
or  of Kaluza-Klein monopoles \cite{Bizon:2006ue} or their  
analogue in nine spacetime dimensions \cite{Bizon:2007zf}. 
    
 It would seem therefore that, just as in four spacetime dimensions,
the natural candidates for solitons in five dimensional supergravity theories
are extreme black holes admitting Killing spinors.
However this is not true. The BPS fuzzball solutions of \cite{Bena:2005va,Berglund:2005vb,Bena:2007kg} 
are, as we shall show in detail in the body of this paper,
complete and non-singular, with no horizons. 
Since a conventional Smarr formulae have been established  for five 
dimensional black holes, at least in a special case \cite{Gauntlett:1998fz},  
and a Lichnerowicz type theorem proved \cite{Shiromizu:2012hb},
the question  naturally arises as to how the fuzzball solutions
get round the these No-Go Theorems. It is the purpose of this paper
to answer this question. In brief, the answer is 
that if one includes both Chern-Simons terms, which were correctly
included in \cite{Gauntlett:1998fz} and takes into account the
possibility that the four-dimensional spatial manifold  
may be topologically non-trivial, and in particular may have
non-trivial second homology group, there is an extra bulk term
in the Smarr formula which is sufficient 
to prevent the arguments in \cite{Breitenlohner:1987dg} being extended   
from four to five, or indeed higher, spacetime dimensions.
To put this in slightly different terms: the combination 
of non-trivial topology and Chern-Simons interactions allows, 
in five  spacetime dimensions,  what is not possible in 
four spacetime dimensions; the concrete relealistion of Wheeler
and Misner's Geometrodynamics  programme in which particles have their
origin in  field lines trapped in the topology of space \cite{Misner}.

\section{Global structure}
\label{GlobalStr}

In this section we shall  consider the problem in greater generality
and treat  $D$-dimensional spacetimes $\{M,g\}$  
whose  metrics  are stationary 
(at least in some neighborhood of infinity) and so admit a Killing vector field $\bK= K^{\mu} \frac{\p} {\p x^\mu }=  \frac{\p}{\p t }$ in adapted local coordinates  $X^\mu= (t,x^i)\,, \mu=0,1,2,\dots, D-1$. In these adapted fibre coordinates the metric takes the form:
\begin{equation}
ds_D  ^2 ~=~ g_{\mu \nu} dx ^\mu dx ^\nu ~=~-  \frac{1}{Z^2(x^j) }  ( dt + k_i(x^k)  dx ^i) ^2 ~+~  \gamma_{ij}(x^k)  dx^i dx ^j
\,. 
\label{Dmetric}
\end{equation} 
{\sl Locally}  we may think of the spacetime manifold $M$ as  an $\BR$-bundle over some $D$-dimensional space of orbits $\cB$  with coordinates $x^i$ and we may think of the hypersurfaces $t={\rm constant}$ as a local section.
The metric $\gamma_ {ij}$ is the projection  of the spacetime metric $g_{\mu \nu}$  orthogonal to  the fibres whose coordinate  is $t$ and the vector field, $k_i$, defines the horizontal, sometimes called in this context
the Sagnac, connection.   This vector field is also sometimes known as the angular momentum vector.   The metric, $\gamma_{ij}$, is Riemiannian    as long as $\frac{\p}{\p t}$ is timelike, 
but is ill-defined at places where   $\frac{\p}{\p t}$ becomes null.  It should be distinguished from the metric induced on the local hypersurfaces $t={\rm constant}$:
\begin{equation}
g_{ij}= \gamma_{ij} - Z^{-2} k_i k_j \,.  
\label{induced1} 
\end{equation}
If we change the  section my means of the coordinate transformation 
\begin{equation}
t ~\rightarrow~  \tilde t =  t + f(x^i) 
\end{equation}
then the induced metric $g_{ij}$ will change since  
\begin{equation}
k_i \rightarrow \tilde k_i  - \frac{\p f}{\p x^i}  \,, 
\end{equation}
but the metric $\gamma_{ij}$  orthogonal to the fibres
will remain unchanged.

In what follows we will explore the global
situation. We have, since $K^\mu =\delta^\mu_0$,  
\begin{equation}
-g_{tt}=-  g_{\mu \nu} K^\mu K^\nu = \frac{1}{Z^2} \,.  
\end{equation}
Thus, if $\frac{1}{Z^2} \ge 0$ everywhere,  then the fibres 
are non-spacelike and become lightlike at points for which $ \frac{1}{Z^2} =0 $. The ``critical''  hypersurface at  which $\frac{1}{Z^2}=0$, may be timelike or null.   The latter represents a degenerate (or ``extreme'') 
Killing horizon, \ie\ a null hypersurface  whose null 
generators coincide with the orbits of the Killing  vector field $\bK$. Since, locally, a future-directed causal (\ie\  timelike or null)  curve may only cross a null hypersurface in one direction, such a surface acts as a stationary one-way membrane.   

Time-like critical hypersurfaces are less familiar. Such a surface is certainly not a one-way membrane since nothing locally  prevents a future directed  causal curve 
crossing a timelike hyperspace in either direction.  If $g_{tt}$ merely
changed sign as one  crossed  a timelike hypersurface, then 
the hypersuface would locally bound a region, called an ergo-region, in which the Killing vector field becomes spacelike, and the surface on which  $\frac{1}{Z^2}=0$ 
would be an an ergo-surface\footnote{Often these surfaces have topology  ${\Bbb R} \times S^{D-2} $ and are then called ergo-spheres.}.       However if  $g_{tt}$ is never  positive, and so 
there is no ergo-region and no generally accepted term
for the surface on which  $\frac{1}{Z^2}=0$. The
occurrence of such surfaces, at which   $\frac{1}{Z^2}=0$
typically has a double zero, often arise in supersymmetric
spacetimes for which $\bK$ has a spinorial square root, that
is,  there exists a spinor field $\epsilon$
for which 
\begin{equation}
K^\mu = \bar \epsilon \gamma ^\mu \epsilon \,. \label{Kspinor}  
\end{equation}
One can show that  the right-hand side of (\ref{Kspinor}) is never spacelike.
If $\frac{1}{Z^2}$ has a double zero, the normal to the
hypersurface $\frac{1}{Z}=0$ will be timelike if
$g^{ij} \frac{\p_i Z \p_j  Z}{Z^4}$ is positive there and
null if $g^{ij} \frac{\p_i Z \p_j  Z}{Z^4}$ vanishes there.     

The section $t={\rm constant}$ has a normal $\p_\mu t$ 
\begin{equation}
g^{tt}= g^{\mu \nu} \p _\mu t \p_\nu t = -
(Z^2 - \gamma ^{i j} k_i k_i)    
\end{equation}
and will be everywhere timelike, and the section everywhere
spacelike if 
\begin{equation}
(Z^2 - \gamma ^{i j} k_i k_i) >0\,.
\label{cond1}\end{equation}

Note that 
\begin{equation}
\det g_{ij} = \frac{1}{Z^2}(\det \gamma_{ij}) \, (Z^2  - \gamma ^{i j} k_i k_i)
\end{equation}
and so a necessary condition that the metric induced on the section
is positive definite is
\begin{equation}
\frac{1}{Z^2} (\det \gamma_{ij}) \, (Z^2  - \gamma ^{i j} k_i k_i) >0\,.
\label{cond2} \end{equation}
If conditions (\ref{cond1}) and (\ref{cond2}) hold globally
then any future directed causal  curve 
may cross a section once and only once and one
 may regard $t$ as a {\it global time function}.  The spacetime is then 
{\it stably causal} \cite{Hawking:1968jt}   
and the  sections $t={\rm constant}$  can be viewed as  {\it global  Cauchy surfaces}.
If $\cB$ is the space of orbits, then the topology of the spacetime
manifold will be a product $M\equiv \BR \times \cB$.  

A simple example of the situation  we are interested in
is the product metric on $AdS_3 \times \frac{1}{4}  S^2$ \cite{Bena:2005va,Berglund:2005vb,Bena:2007kg}
\begin{equation}
ds_5^2 = - \cosh ^2 \xi d \tau ^2 + d \xi ^2 + \sinh ^2 \xi d\varphi_1^2
+ \frac{1}{4} \bigl( d \theta ^2 + \sin ^2 \theta d \varphi_2^2  \bigr )    
\label{ads3s2}
\end{equation}
with the  Killing vector
\begin{equation}
\bK = \frac{\p}{\p \tau} -\frac{\p}{\p \varphi_1} + 2 
\frac{\p}{\p \varphi_2} 
\end{equation}
for which
\begin{equation}
g(K,K) = - \cos ^2 \theta. 
\end{equation}
One may think of the integral curves of $\bK$ as  world-lines of  non-space-like ``observers'' that are time-like everywhere except for $\theta =   \frac{\pi}{2}$.
The metric is clearly geodesically  complete and non-singular
and the Killing vector field  $\bK$ regular and timelike
everywhere except on the equator of the $S^2$, $\theta = \frac{\pi}{2}$.
If we set $\theta=\frac{\pi}{2}$ in the metric (\ref{ads3s2}) we get a
regular four-dimensional metric of signature $-+++$, that is 
$\theta = \frac{\pi}{2}$ is a timelike hypersurface. 
While this example may seem a little contrived, we will see in Section \ref{AdSS} that it is, in fact, an archetypical model for the local description of critical surfaces in a fuzzball model.

\section{Some Microstate geometries}

\subsection{The $\Neql{2}$ supergravity theory}

The simplest  candidate  microstate geometries have been  constructed in   $\Neql{2}$,  five-dimensional supergravity coupled to two vector multiplets.   Including the gravi-photon there are three vector fields and  two independent scalars which may   conveniently be parametrized by   the fields, $X^I$, $I=1,2,3$ satisfying the constraint $X^1 X^2 X^3  = 1$. 
The bosonic action is 
\begin{eqnarray}
  S ~=~ \int\!\sqrt{-g}\,d^5x \Big( R  -\coeff{1}{2} Q_{IJ} F_{\mu \nu}^I   F^{J \mu \nu} - Q_{IJ} \partial_\mu X^I  \partial^\mu X^J -\coeff {1}{24} C_{IJK} F^I_{ \mu \nu} F^J_{\rho\sigma} A^K_{\lambda} \bar\epsilon^{\mu\nu\rho\sigma\lambda}\Big) \,,
  \label{5daction}
\end{eqnarray}
with $I, J =1,2,3$. The metric for the kinetic terms is 
\begin{equation}
 Q_{IJ} ~=~    \frac{1}{2} \,{\rm diag}\,\big((X^1)^{-2} , (X^2)^{-2},(X^3)^{-2} \big) \,.
\label{scalarkinterm}
\end{equation}

We are interested in  five-dimensional stationary space-times, $\cM_5$,
whose Lorentzian metric may be cast in the  local form:
\begin{equation}
ds_5^2 ~=~ -Z^{-2} \,(dt + k)^2 ~+~ Z \, ds_4^2  \,, 
\label{metform}
\end{equation}
where, compared to (\ref{Dmetric}), it is convenient to introduce the warp factor, $Z$, in front of the general metric, $ds_4^2$ on the four-dimensional base manifold,  $\cB$.  The metric induced on the local
hypersurfaces $t={\rm constant}$ is now given by:
\begin{equation}
ds_{\rm induced}^2 ~=~ -Z^{-2} \, k^2 ~+~ Z \, ds_4^2  \,.
\label{induced2}
\end{equation}

The matter fields are assumed to be time-independent
and therefore the Maxwell fields may be decomposed into 
electric and magnetic components:
\begin{equation}
A^I   ~=~  -  Z_I^{-1}\, (dt +k) ~+~ B^{(I)}  \,,
\label{Aform}
\end{equation}
where $B^{(I)}$ is a one-form on $\cB$.   It will prove  convenient to define magnetic field strengths:
\begin{equation}
\Theta^{(I)}    ~\equiv~  d B^{(I)}    \,.
\label{Thetadefn}
\end{equation}
%

\subsection{BPS  solutions}

The solutions of \cite{Gauntlett:2002nw,Gauntlett:2004qy,Bena:2005va,Berglund:2005vb,Bena:2007kg} are  BPS, {\it i.e.}  supersymmetric and they are  be obtained by  requiring that they admit Killing spinor fields.  Specifically, one  seeks solutions that preserve exactly the same supersymmetries as a BPS black hole with the same electric charges.  This leads to what later became known as the ``floating brane Ansatz'' \cite{Bena:2009fi} because the constituent charges are carried by branes and those branes must obey a `zero-force' condition in a BPS solution.  This means that the scalars and warp factors are  related to the electric potentials  via:
\begin{equation}
Z ~\equiv~ \big( Z_1 \, Z_2 \, Z_3  \big)^{1/3}\,,\quad    X^1    =\bigg( \frac{Z_2 \, Z_3}{Z_1^2} \bigg)^{1/3} \,, \quad X^2    = \bigg( \frac{Z_1 \, Z_3}{Z_2^2} \bigg)^{1/3} \,,\quad X^3   =\bigg( \frac{Z_1 \, Z_2}{Z_3^2} \bigg)^{1/3}  \,.
\label{XZrelns}
\end{equation}
The conformally rescaled base metric, $ds_4^2$, is then 
required to be hyper-K\"ahler and supersymmetric configurations are  
obtained by solving the system of equations \cite{Bena:2004de,Bena:2005va,Berglund:2005vb}:
\begin{eqnarray}
 \Theta^{(I)}  &~=~&  \star_4 \, \Theta^{(I)} \label{BPSeqn:1} \,, \\
 \nabla^2  Z_I &~=~&  {1 \over 2  }  C_{IJK} \star_4 (\Theta^{(J)} \wedge
\Theta^{(K)})  \label{BPSeqn:2} \,, \\
 dk ~+~  \star_4 dk &~=~&  Z_I \,  \Theta^{(I)}\,,
\label{BPSeqn:3}
\end{eqnarray}
where $\star_4$ is the Hodge dual taken with respect to the 
four-dimensional metric, $ds_4^2$,  and the structure constants are given by $C_{IJK} ~\equiv~ |\epsilon_{IJK}|$.  More generally, when the supergravity is coupled to more $\Neql{2}$ vector multiplets, these structure constants are precisely those that determine the structure of the vector multiplet sector and its scalar coset.

For the metric (\ref{metform}) to be asymptotically flat and  the vector kinetic term in  (\ref{5daction}) to be well-behaved at infinity one usually requires that $Z_I $ goes to a non-zero constant at infinity.  By rescaling coordinates and fields one can, without loss of generality, take 
\begin{equation}
Z_I ~\to~ 1 
\label{Zasymp}
\end{equation} 
at infinity.  This will then give the vector kinetic term its canonical normalization at infinity. As we will see in Section \ref{AdSS}, if one wants different asymptotics then one does not necessarily impose  (\ref{Zasymp}).

\section{Smarr Formula in $4+1$ spacetime dimensions}
\label{Smarr1}

In many circumstances, Smarr's formula enables one to relate the mass of a solution to properties on  interior boundaries.  Moreover, if there are no such boundaries because the solution is smooth and horizonless, one typically finds that the mass must be zero.  Thus Smarr's formula lies at the heart of the belief that there are ``No solitons without horizons.''  We will show how solutions arising from the action   (\ref{5daction}) avoid this conclusion precisely because of the Chern-Simons term.   Interestingly enough, the role of Chern-Simons terms has been carefully analysed in the context of horizon topologies  \cite{Hollands:2012cc,Hollands:2012xy} but the consequences of topological Chern-Simons contributions in the bulk space-time do not appear to have been considered to date.

\subsection{Equations of Motion}

The Einstein equations coming from   (\ref{5daction}) are:
\begin{equation}
R_{\mu \nu} -\coeff{1}{2} g_{\mu \nu} R  ~=~   Q_{IJ}\,\Big[  F^{I}_{\,\mu \rho}  \, {{F^J}_\nu \,}^\rho   - \coeff{1}{4} \, g_{\mu \nu}  \,  F^{I}_{\, \rho \sigma} F^{ J\, \rho \sigma} +  \partial_\mu X^I \,  \partial_\nu X^J  
 -  \coeff{1}{2} \, g_{\mu \nu}  \, g^{\rho \sigma}  \,  \partial_\rho  X^I \,  \partial_\sigma X^J   \Big]\,.  \label{Einstein1} 
\end{equation}
Taking traces and rearranging gives the equation:
\begin{equation}
R_{\mu \nu} ~=~  Q_{IJ}\,\Big[  F^{I}_{\, \mu \rho}  \, {{F^J}_\nu \,}^\rho   - \coeff{1}{6} \, g_{\mu \nu}  \,  F^{I}_{\, \rho \sigma} F^{ J\, \rho \sigma} +  \partial_\mu X^I \,  \partial_\nu X^J  
 \Big]\,.  \label{Einstein2} 
\end{equation}

The Maxwell equations coming from    (\ref{5daction})  are:  
\begin{equation}
\nabla_{\rho} \big(Q_{IJ}  {{F^J}^{\rho}}_\mu \big)  ~=~  J^{CS}_{I \,\mu}  \,, \label{Max1} 
\end{equation}
where the Chern-Simons currents are given by:
\begin{equation}
J^{CS}_{I \,\mu}  ~\equiv~  \coeff {1}{16} \, C_{IJK}\, \epsilon_{\mu \alpha \beta \gamma \delta } \, F^{J\, \alpha \beta}  \, F^{k\,\gamma \delta}   \,. \label{CScurrents} 
\end{equation}

Define dual $3$-froms, $G$, by 
\begin{equation}
G_{I\, \rho \mu \nu}  ~\equiv~  \coeff {1}{2} \, Q_{IJ} \, F^{J\, \alpha \beta}   \, \epsilon_{\alpha \beta \rho \mu  \nu  }     \label{Gdefns} 
\end{equation}
and introduce the inverse, $Q^{IJ}$ of $Q_{IJ}$:
\begin{equation}
Q^{IJ}   \, Q_{JK} ~= ~ \delta^I_K  \,. \label{QIJinv} 
\end{equation}
If follows from the Bianchi identities for $ F^{J}_{\mu  \nu}$ that $G_{J}$ satisfies:
\begin{equation}
\nabla_{\rho} \big(Q^{IJ} {G_{J}}^{\,\mu  \nu \rho  }   \big)  ~=~ 0 \,, \label{MaxG} 
\end{equation}
Similarly, from the equations of motion (\ref{Max1}) for $ F^{J}_{\mu  \nu}$  one has 
\begin{equation}
\nabla_{[ \lambda}  G_{| J | \, \rho \mu  \nu  ] }  ~=~ + \coeff{3}{8} \, C_{IJK} \,  F^{J}_{[\lambda \rho} \, F^{K}_{\mu  \nu]} \qquad \Leftrightarrow \qquad d G_I ~=~ + \coeff{1}{4} \, C_{IJK} \,  F^{J} \wedge F^{K}  \,.\label{BianchiG} 
\end{equation}
where $| J |$ means that the index $J$ is not involved in the skew-symmetrization bracket $[\dots]$.

One can  easily verify that
\begin{equation}
Q^{IJ}   \, G_{I\, \mu \rho  \sigma} \,{G_{J}}^{\nu \rho  \sigma}     ~=~  Q_{IJ} \, \big( 2\, F^{I}_{\, \mu  \rho} \,    F^{J\,  \nu  \rho}   -   \delta_\mu^\nu \, F^{I}_{ \, \rho  \sigma}   \,   F^{J \,\rho  \sigma}    \big)      \label{GGFF} 
\end{equation}
and so we may rewrite the Einstein equation (\ref{Einstein2}) as
\begin{equation}
R_{\mu \nu} ~=~  Q_{IJ}\,\Big[  \coeff{2}{3}\, F^{I}_{\, \mu \rho}  \, {{F^J}_\nu \,}^\rho  +  \partial_\mu X^I \,  \partial_\nu X^J  
 \Big] ~+~  \coeff{1}{6}\, Q^{IJ} \,  G_{I\, \mu \rho  \sigma} \,{G_{J\, \nu}}^{ \rho  \sigma}       \,.  \label{Einstein3} 
\end{equation}
%

\subsection{Invariances}

As remarked upon above, we shall make the assumption
 that the matter fields share the  symmetry of the metric.
In particular we assume that they are invariant under diffeomorphisms 
generated by the Killing vector, $K^\mu$: 
\begin{equation}
{\cal L}_K  F^I ~=~0 \,, \qquad   {\cal L}_K  G_I ~=~0 \,, \qquad {\cal L}_K  X^I ~=~0  \,,   \label{invariances1}  
\end{equation}
where $\cL_K$ denotes the Lie derivative. A formula of Cartan states that for a $p$-form, $\alpha$, one has
\begin{equation}
{\cal L}_K  \alpha ~=~d (i_K( \alpha) ) ~+~  i_K(d \alpha)  \,.  \label{Cartan}  
\end{equation}
Taking $ \alpha  =  F^I $ we have, locally, 
\begin{equation}
 K^\rho  F^I_{ \rho \mu} ~=~  \partial_\mu  \lambda^I  \,, \label{KdotF}  
\end{equation}
for some functions $ \lambda^I$.

If the space-time manifold were not simply connected one could, in principle, encounter jumps in value of $\lambda^I$ is one were to integrate (\ref{KdotF}) around a closed curve.  To avoid this issue we shall, from now on, assume that our space-time manifold, ${\cM_5}$ is simply connected.  With this assumption, the arbitrary constants in the definitions of the functions, $\lambda^I$, may be fixed by requiring that the $\lambda^I$ vanish at infinity.  As we will see, in Section \ref{Smarr2}, this choice of boundary condition will crucially affect the the details of the Smarr formula.  Physically, the functions, $\lambda^I$, are magnetostatic potentials of the $3$-forms, $G_I$, or, equivalently, electrostatic potentials of the $2$-forms, $F^I$.   

Taking $ \alpha  =  G_I$ we have 
\begin{align}
d(i_K (G_I)) ~=~ &   -i_K (d G_I) ~=~   -\coeff{1}{4}  \, C_{ILM} \, i_K (  F^{L}\wedge F^{M})   ~=~   -\coeff{1}{2}  \, C_{ILM} \,  d \lambda^L \wedge  F^{M}  \nonumber  \\ ~=~ &   -\coeff{1}{2}  \,C_{ILM} \,  d( \lambda^L \, F^{M} )   \label{dKdotG}  
\end{align}
where we have used (\ref{BianchiG}) and  (\ref{KdotF}). While the assumption of simple connectivity is a weak one because we could pass to a covering space, we cannot assume that the $H^2(\cM_5)$ is trivial. Indeed, this is the crucial issue that makes solitons possible.  Thus we deduce that 
\begin{equation}
K^\rho  G_{I\,  \rho \mu \nu}  ~=~   \partial_\mu  \Lambda_{I\, \nu} - \partial_\nu  \Lambda_{I\, \mu} ~-~ \coeff{1}{2} \, C_{IJK}  \, \lambda^J  F^{K}_{\mu  \nu} ~+~ H_{I \, \mu \nu}\,,   \label{KdotG}  
\end{equation}
where $\Lambda_I$ are globally defined one-forms and $H_I$ are closed but not exact two forms. That is, we cannot write $H_I = d \nu_I$ where $\nu_I$ are globally well-defined one-forms.

Using (\ref{KdotF}) and (\ref{KdotG}) we see that 
\begin{align}
 K^\mu \big( Q_{IJ}\,   F^{I}_{\, \mu \rho}  \, {{F^J}_\nu \,}^\rho  \big) ~=~&   - \nabla_\rho \big(  Q_{IJ}\, \lambda^I \, {F^J}^{ \rho\nu}  \big) ~+~   \coeff{1}{16} \, C_{IJK}  \,  \epsilon^{\nu \alpha \beta \gamma \delta} \, \lambda^I \, {F^J}_{  \alpha \beta} \,{F^K}_{\gamma \delta} \\
  K^\mu \big( Q^{IJ}\, G_{I \, \mu  \rho \sigma}  \, {G_J}^{\, \nu  \rho \sigma}  \big) ~=~&  - 2\, \nabla_\rho \big(  Q^{IJ}\, \Lambda_{I \, \sigma} \, {G_J}^{\,   \rho \nu \sigma} \big) ~-~   \coeff{1}{4} \, C_{IJK}  \,  \epsilon^{\nu \alpha \beta \gamma \delta} \, \lambda^I \, {F^J}_{  \alpha \beta} \,{F^K}_{\gamma \delta} \nonumber \\
  &  ~+~  Q^{IJ}\, H_I^{\rho \sigma} \, {G_J}_{\rho \sigma \nu }
\end{align} 
and hence,  Einstein's equations (\ref{Einstein3}) become:
\begin{equation}
 K^\mu R_{\mu \nu} ~=~   - \coeff{1}{3}\, \nabla^\mu \, \big[ \, 2\,  Q_{IJ}\, \lambda^I \, {F^J}_{ \mu\nu}  
 ~+~   Q^{IJ}\, {\Lambda_{I }}^{\sigma} \, {G_J}_{\,   \mu \nu \sigma} \, \big] ~+~ \coeff{1}{6}\,  Q^{IJ}\, H_I^{\rho \sigma} \, {G_J}_{\rho \sigma \nu } \,,  \label{KdotR} 
\end{equation}
where we have used $K^\mu \partial_\mu X^I =  \cL_K X^I  =0$.  Note that the $\lambda (* F\wedge F)$  terms  have canceled in (\ref{KdotR}).  
The last term in  (\ref{KdotR}) may be expressed as 
\begin{equation}
\coeff{1}{6}\,  Q^{IJ}\, H_I^{\rho \sigma} \, {G_J}_{\rho \sigma \nu }  ~=~  \coeff {1}{12} \, \epsilon_{\alpha \beta \rho \sigma  \nu  }   F^{I\, \alpha \beta}\,H_I^{\rho \sigma}       \,.  \label{HdotG} 
\end{equation}

A similar result was obtained in \cite{Gauntlett:1998fz}, where it was also noted that precisely in five dimensions the $\lambda (* F\wedge F)$  cancel out in the expression (\ref{KdotR}).  However, they assumed the global existence of the vector potentials, $A^I$. In the ``no-go'' theorem of \cite{Shiromizu:2012hb} there were no Chern-Simons terms but in their equations (37) and (38) they assumed the existence of global potentials, $\Psi$ and $\Phi$, and a crucial term, analogous to (\ref{HdotG}) is missing in their equation (48).

\subsection{Mass and Charge}

\subsubsection{Expansions at infinity}
\label{Normalizations}

To get the correctly normalized asymptotic charges for an asymptotically flat metric in a $D$-dimensional space-time one should start from the canonically normalized action:
\begin{equation}
S ~=~  \int d^D x \, \sqrt{-g} \,  \bigg( \frac{R}{16\pi G_D}  ~+~{\cal{L}}_{\rm matter} \bigg) \,,
\label{canonact}
\end{equation}
where $G_D$ is the Newton constant.   The Einstein equations are, as usual, 
\begin{equation}
R_{\mu \nu}  - \frac{1}{2} \, R \, g_{\mu \nu} ~=~  8 \pi G_D  \, T_{\mu \nu} \,,
\label{Ein1}
\end{equation}
where $T_{\mu \nu}$ is the canonically normalized energy-momentum tensor.  The Einstein equations may be rewritten as   
\begin{equation}
R_{\mu \nu}  ~=~  8 \pi G_D \,  \Big(  T_{\mu \nu} - \frac{1}{(D-2)} \,  T \, g_{\mu \nu} \Big) \,,
\label{Ein2}
\end{equation}
where $T$ is the trace of $T_{\mu \nu} $.

If one linearizes around a flat metric one can then define the momentum and angular momentum of the configuration by integrating over a space-like hyper surface, $\Sigma$: 
\begin{equation}
P^\mu  ~=~  \int_\Sigma d^{D-1} x  \, T^{\mu 0}  \,,  \qquad 
J^{\mu \nu}  ~=~     \int_\Sigma d^{D-1} x  \, \big( x^\mu  T^{\nu 0} - x^\nu  T^{\mu 0} \big) \,.
\end{equation} 
One can then use the linearized Einstein equations to show that in a rest frame one has  \cite{Gibbons:1993xt,Sabra:1997yd,Myers:1986un, Peet:2000hn}:
\begin{eqnarray}
g_{00} &=&  -1 ~+~  \frac{16\pi G_D} {(D-2)\, A_{D-2}}  \frac{M}{\rho^{D-3} }~+~ \dots  \,,  \label{asympg1} \\
g_{ij}   &=&  1  ~+~  \frac{16\pi G_D} {(D-2)\, (D-3)\, A_{D-2}}   \frac{M}{\rho^{D-3}} ~+~ \dots  \,,\label{asympg2} \\ 
g_{0i} &=&   \frac{16\pi G_D} { A_{D-2}}   \frac{x^j J^{ji} }{\rho^{D-1}}~+~ \dots  \label{asympg3}\,,
\end{eqnarray}
where $\rho$ is the radial coordinate and $A_{D-2}$ is the volume of a unit $(D-2)$ sphere.  In particular, one has $A_3 = 2 \pi^2$.

Note that in the linearized system it follows from   (\ref{Ein2}) that 
\begin{equation}
R_{00} ~\approx~   8 \pi G_D \,  \Big( T_{00} - \frac{1}{(D-2)} \, g^{00} \,  T_{00}  \, g_{00}\Big)   ~=~ 8 \pi G_D \,  \frac{(D-3)}{(D-2)}  \,  T_{00} \,.
\label{normR00}
\end{equation}

For the Maxwell action 
\begin{equation}
S_{\rm Maxwell} ~=~ \int d^D x \, \sqrt{-g} \,  \Big(-  \frac{1}{4} \, F_{\mu \nu} F^{\mu \nu}  \Big) \,,
\label{Maxact}
\end{equation}
 the asymptotic electric charge is given by the expansion:
\begin{equation}
F_{0 \rho}  ~=~    (D-3)\, \frac{Q}{\rho^{D-2} }  \,.
\label{asympQ}
\end{equation}
Note that the more standard normalization of charge that is adapted to Gaussian integrals would have a different overall  factor involving $A_{(D-2)}^{-1}$ but here we have chosen to normalize the $U(1)$ charges so as to be consistent with the usual literature on bubbled geometries (see, for example, \cite{Bena:2007kg}).

More generally, the expansions (\ref{asympg1})--(\ref{asympg3}) and (\ref{asympQ}) may be used to define asymptotic charges of a generic, asymptotically-flat metric.

\subsubsection{Normalizing the Komar Integrals}
\label{NormKomar}

If the metric has Killing vectors then they can be used to define a globally conserved quantity via a Komar integral.   
Indeed, if  $K$ is a time-like Killing vector then the following Komar integral defines a conserved mass:
\begin{equation}
 \int _{S^{D-2}}   \, * d K~=~   \int _{S^{D-2}}  \,\big(\partial_\mu K_\nu - \partial_\nu K_ \mu  \big)  d \Sigma^{\mu \nu} ~ \,.
 \label{Komar1}
\end{equation}
where $S^{D-2}$ is a closed, space-like surface at infinity.  If  the  configuration is smooth on a space-like hypersurface, $\Sigma$, then one can write  
\begin{equation}
 \int _{S^{D-2}}   \, * d K~=~    \int _{\Sigma}   \, d* d K    ~=~   -2 \,  \int _{\Sigma}   \, *(K^\mu R_{\mu \nu} dx^\nu)    \,.
 \label{Komar2}
\end{equation}
where we have used the fact that for a Killing vector $\nabla^2 K^\mu ~=~ -  R_{\mu \nu}K^\nu$. 

To normalize the Komar mass one can consider, once again, a linearization and use (\ref{normR00}) to obtain:
\begin{equation}
M ~=~  -  \frac{1}{16\pi G_D} \,  \frac{(D-2)}{(D-3)}  \,   \int _{S^{D-2}}   \, * d K   ~=~  -  \frac{1}{16\pi G_D} \, \frac{(D-2)}{(D-3)}  \,    \int _{S^{D-2}}  \,\big(\partial_\mu K_\nu - \partial_\nu K_ \mu  \big)  d \Sigma^{\mu \nu} ~ \,.
 \label{normKomar}
\end{equation}
One may check this normalization against that of (\ref{asympg1}) but observing that to leading order at infinity, $K = g_{00} dt$ and hence 
\begin{equation}
* dK ~\approx~  -  (\partial_\rho g_{00}) \, *(dt \wedge d \rho)  ~\approx~ - \frac{16\pi G_D \, (D-3)} {(D-2)\, A_{D-2}}  \frac{M}{\rho^{D-2} }  \, d{\rm vol}({S^{D-2}}) \,,
 \label{approxKomar}
\end{equation}
which is consistent with (\ref{normKomar}).

\subsubsection{Komar Integrals in five dimensional supergravity}
\label{KomarFiveD}

We now specialize to our five-dimensional theory and take as our starting point the Komar integral for the ADM mass:
\begin{equation}
\frac{16 \pi G_5}{3} \, M ~=~  - \frac{1}{2}\,  \int _{\infty}    \big(\partial_\mu K_\nu - \partial_\nu K_ \mu  \big)  d \Sigma^{\mu \nu}   \,,
 \label{ADMmass}
\end{equation}
where the integral is taken over an $S^3$ at spatial infinity and $d \Sigma^{\mu \nu}$ is the volume form on this sphere.  

Once again we suppose that there is a Cauchy surface,  $\Sigma$, whose boundary at infinity is the $S^3$ above but now we will allow it to also have interior boundaries,  $\partial \Sigma_{int}$.  Then we find:
\begin{align}
\frac{16 \pi G_5}{3} \, M ~=~ &   \int _{\Sigma}  R_{\mu  \nu}  K^\mu\,  d \Sigma^{\nu}   ~+~  \frac{1}{2}\,  \int _{\partial \Sigma_{int}}    \big(\partial_\mu K_\nu - \partial_\nu K_ \mu  \big)  d \Sigma^{\mu \nu}     \label{Mform1}     \\
~=~  & \int _{ \Sigma}   \Big[  \coeff{1}{6}\, Q^{IJ}\, H_{I \rho \sigma} \, {G_J}^{\rho \sigma \nu } ~-~ \coeff{1}{3}\,  \nabla_\mu \Big( 2\, Q_{IJ}\, \lambda^I  {F^J}^{ \mu \nu}  
+  Q^{IJ}\, {\Lambda_{I}}_{\sigma} {G_J}^{\,  \sigma \mu \nu }  \Big)  \Big]  \, d \Sigma_{\nu}  \label{Mform1b}    \\ 
 &~+~  \frac{1}{2}\,  \int _{\partial \Sigma_{int}}    \big(\partial_\mu K_\nu - \partial_\nu K_ \mu  \big)  d \Sigma^{\mu \nu}  \label{Mform2}       \,,
\end{align}
where we have used  (\ref{KdotR}).  If we assume that the second term on the right-hand side of (\ref{Mform1b}) falls off sufficiently fast at infinity then we may write the mass as 
\begin{align}
\frac{16 \pi G_5}{3} \, M ~=~ &\coeff{1}{6}\,  \int _{ \Sigma}   \Big[  Q^{IJ}\, H_{I \rho \sigma} \, {G_J}^{\rho \sigma \nu }\Big]  \, d \Sigma_{\nu}   \nonumber \\
 &~+~  \int _{\partial \Sigma_{int}}  \, \Big[ -\coeff{1}{3}\, \Big( 2\, Q_{IJ}\, \lambda^I  {F^J}^{ \mu \nu}  
+  Q^{IJ}\, {\Lambda_{I}}_{\sigma} {G_J}^{\,  \sigma \mu \nu }  \Big) ~+~ \coeff{1}{2}\,   \big(\partial_\mu K_\nu - \partial_\nu K_ \mu  \big)  \Big] d \Sigma^{\mu \nu} \,.\label{ADMmassres} 
\end{align}

In the standard applications of the Smarr formula (see, for example, \cite{Gauntlett:1998fz,Shiromizu:2012hb}), the two-form,  $H_I$, is assumed to be zero and the interior boundaries are horizons. Thus  the boundary terms relate the ADM mass to horizon areas,  charges and angular momenta.  If there are no horizons and one assumes that $H_I = 0$ then the ADM mass vanishes.  However,  using (\ref{Mform1}), we have 
\begin{equation}
\frac{16 \pi G_5}{3} \, M ~=~    \int _{ \Sigma}\, R_{00}  \, d^4 x  \,.\label{posmass} 
\end{equation}
From (\ref{Einstein3})  we see that $R_{00} \ge 0$ and vanishes if and only if $F^I_{0j} = 0$, $\partial_t  X^I = 0$ and $G_{0ij} = 0$.  Since $G_I$ is the dual of $F^I$, this means that $F^I_{ij} = 0$ and hence $F^I$ and $G_I$ vanish  identically.  The scalars therefore have no source and if one parameterizes them with $X^1= e^{\phi_1+ \phi_2}$, $X^1= e^{\phi_1- \phi_2}$ and $X^3= e^{ - 2\, \phi_1}$ then the $\phi_a$ must be harmonic.  Since $\partial_t  \phi_a = 0$ there are no non-trivial solutions that are smooth and fall off at infinity.  Hence we must have $ \phi_a =0$ and $X^I =1$ everywhere. Thus the complete solution would necessarily be trivial. This result is consistent with \cite{Shiromizu:2012hb}, which analyses a simpler theory with a single $3$-form with no Chern-Simons term and makes the assumption that the potentials are globally well-defined.
 
On the other hand, if $H_I$ is not zero and there are no inner boundaries, we conclude that  $M$ is not only positive but must be given by:
\begin{equation}
M  ~=~     \frac{1}{32 \pi G_5} \,  \int _{ \Sigma}   \Big[  Q^{IJ}\, H_{I \rho \sigma} \, {G_J}^{\rho \sigma \nu }\Big]  \, d \Sigma_{\nu}    \label{Manswer} \,.
\end{equation}
 In the corresponding situation in four space-time dimensions there is no analogue of the $H_I$ and one concludes \cite{Breitenlohner:1987dg} that for ungauged  supergravity theories in general there are no soliton solutions unless horizons are present. In five dimensions however, we have seen that provided our space-time is topologically non-trivial, with non-vanishing $H^2(\cM_5)$, there is no obstacle to regular solitons states without horizons.

\section{A class of examples}

\subsection{The hyper-K\"ahler base}

Perhaps the simplest hyper-K\"ahler metrics are those based upon a $U(1)$ fibration over a flat $\IR^3$:
\begin{equation}
ds_4^2 ~=~  h_{ij}dx^i dx^j  ~=~    \frac{1}{V(y^a)} ( d \psi + A_a(y^a) dy^a ) ^2 + V (y^a) dy^a dy ^a   \,,
\label{GHmet}
\end{equation}
where $a=1,2,3$ and
\begin{equation}
\vec \nabla \times  \vec A ~=~  \vec \nabla V 
\label{VAreln}
\end{equation}
where $\vec \nabla$ denotes the standard gradient operator, $\frac{\partial}{\partial y^a}$, on flat, Euclidean $\IR^3$.  Note that (\ref{VAreln}) implies that  $V$ is a harmonic function on $\IR^3$. 
The circle action $ \bL = \frac{\p }{\p \psi}$ generated  is  tri-holomorphic and the triple of functions, $y^a$, are the moment maps for the tri-holomorphic circle action generated by $\bL$ .  

The harmonic functions, $V$, that we will consider take the form
\begin{equation}
V = \varepsilon_0 + \sum_{j=1}^N  \frac{q_j}{ |\vec y  - \vec y^{(j)} | } \label{pot} 
\end{equation}
where $\varepsilon_0$ may be zero or one, and the $q_j$ may be plus or minus one.  Thus, in general $V$ will not be positive, and the metric signature will change from $+4$ to $-4$ when $V$ changes sign.   

However we can always find solutions in which $Z V$ is globally positive:  
\begin{equation}
Z\, V >0  \,, 
\end{equation}
and smooth, except possibly at the points $\vec y^{(j)}$, at which $Z$ is finite and hence, near $\vec y^{(j)}$, one has
\begin{equation}
Z\, V  ~\sim~  \frac{z_j  q_j }{ |\vec y  - \vec y^{(j)} |}  \label{ZV}  \,, 
\end{equation}
for some constants, $z_j$.  If $|q_j| =1$, then the apparent singularity of the metric at $\vec y^{(j)}$  may be eliminated by making the identification
\begin{equation}
0\le \psi \le 4 \pi.
\end{equation}
and changing variables so that $|\vec y  - \vec y^{(j)} | = \frac{1}{4} R^2$.   It follows   that the $4$-metric 
\begin{equation}
Z  ds_4^2  ~=~   \frac{Z}{V } ( d \psi + \vec A \cdot d \vec y  ) ^2 ~+~ Z\, V \, d \vec y   \cdot d \vec y  \,, \label{4metric}
\end{equation}
is smooth and Riemannian in neighborhoods of the points $ \vec y^{(j)}$, provided that $|q_j| =1$.  If $|q_j| \in \ZZ_+$ then the manifold has a simple $\ZZ_{|q_j|}$ orbifold singularity at $\vec y^{(j)}$.  Indeed, away from the locus $V=0$, the metric (\ref{4metric}) non-singular and Riemannian. However, the metric (\ref{4metric})  in the direction of the circle fibre is singular at  $V=0$.   We will refer to the surfaces defined by $V=0$ as {\it critical surfaces} and in Section \ref{CritReg}, we will describe how this singularity at $V=0$ is cancelled in the full five-dimensional metric, (\ref{metform}),  by terms coming from the angular momentum vector, $k$.

The behaviour of the metric at infinity is determined by the asymptotic behaviour of $V$ in (\ref{pot}) and so we  define the parameter, $q_0$, by 
\begin{equation}
q_0  ~\equiv~  \sum_{j=1}^N\, q_j \,.
\label{q0defn}
\end{equation}
If $\varepsilon_0 \ne 0$ then  (\ref{GHmet}) is asymptotic to the flat metric on $\IR^3 \times S^1$ but if $\varepsilon_0 = 0$ then the metric (\ref{GHmet}) is asymptotic to flat $\IR^4$ if and only if $q_0 = +1$.

\subsection{The complete solution}

We now describe the $\psi$-independent solutions  \cite{Gauntlett:2004qy,Bena:2005ni, Bena:2005va,Berglund:2005vb, Bena:2007kg} of the BPS equations (\ref{BPSeqn:1})--(\ref{BPSeqn:3}) for the base metric (\ref{GHmet}).  Introduce a set of frames
\begin{equation}
\hat e^1~=~ V^{-{1\over 2}}\, (d\psi ~+~ A) \,,
\qquad \hat e^{a+1} ~=~ V^{1\over 2}\, dy^a \,, \quad a=1,2,3 \,,
\label{GHframes}
\end{equation}
and two associated sets of two-forms:
\begin{equation}
\Omega_\pm^{(a)} ~\equiv~ \hat e^1  \wedge \hat
e^{a+1} ~\pm~ \coeff{1}{2}\, \epsilon_{abc}\,\hat e^{b+1}  \wedge
\hat e^{c+1} \,, \qquad a =1,2,3\,.\
\label{twoforms}
\end{equation}
The two-forms, $\Omega_-^{(a)}$, are anti-self-dual,  harmonic and
non-normalizable  and they define the
hyper-K\"ahler  structure on the base.  The forms, $\Omega_+^{(a)}$, are
self-dual and can be used to construct harmonic fluxes that are dual to the
two-cycles.  The Maxwell fields, $\Theta^{(I)}$, are required by (\ref{BPSeqn:1}) to self-dual and have the form
\begin{equation}
\Theta^{(I)} ~ \equiv~\sum_{a=1}^3 \, \big(\partial_a \big( V^{-1}\, K^I  \big)\big) \,
\Omega_+^{(a)} \,.
\label{harmtwoform}
\end{equation}
where the $K^I $ are harmonic in $\IR^3$, {\it i.e.}  $\nabla^2 K^I =0$.
It is straightforward to find a {\it local} potential  such that $\Theta^{(I)}  = dB^I$:
\begin{equation}
B^I ~\equiv~  V^{-1}\,   K^I  \, (d\psi ~+~ A) ~+~  \vec{\xi}^I \cdot d \vec y \,,
\label{Bpot}
\end{equation}
where
\begin{equation}
\vec  \nabla \times  \vec{\xi}^I  ~=~ - \vec \nabla K^I \,.
\label{xidefn}
\end{equation}

The solution to (\ref{BPSeqn:2}) for $Z_I$ is 
\begin{equation}
Z_I ~=~ \coeff{1}{2}  \, C_{IJK} \, V^{-1}\,K^J K^K  ~+~ L_I \,,
\label{ZIform}
\end{equation}
where  $C_{IJK} ~\equiv~ |\epsilon_{IJK}|$ and the $L_I$ are three more independent harmonic functions.
If one writes the angular-momentum vector,   $k$, as: 
\begin{equation}
k ~=~ \mu\, ( d\psi + A   ) ~+~ \omega \,, 
\label{kansatz}
\end{equation}
then the solution to (\ref{BPSeqn:3}) is given by:
\begin{equation}
\mu ~=~ \coeff {1}{6} \, C_{IJK}\,  {K^I K^J K^K \over V^2} ~+~
{1 \over 2 \,V} \, K^I L_I ~+~  M\,,
\label{mures}
\end{equation}
where $M$ is yet another harmonic function\footnote{In this section we are using the standard notation for describing these solutions and one should note, in particular, that $M$ is {\it not} the mass of the solution.}  on $\IR^3$. 
One also has
\begin{equation}
\vec \nabla \times \vec \omega ~=~  V \vec \nabla M ~-~
M \vec \nabla V ~+~   \coeff{1}{2}\, (K^I  \vec\nabla L_I - L_I \vec
\nabla K^I )\,.
\label{omegeqn}
\end{equation}
The obvious multi-centre solution has:
\begin{equation}
 V = \varepsilon_0 ~+~ \sum_{j=1}^N \,  {q_j  \over r_j} \,, \qquad K^I ~=~ k^I_0 ~+~  \sum_{j=1}^N \, {k_j^I \over r_j} \,,
\label{KVform}
\end{equation}
\begin{equation}
 L^I ~=~ \ell^I_0 ~+~  \sum_{j=1}^N \, {\ell_j^I \over r_j} \,, \qquad
M ~=~ m_0 ~+~  \sum_{j=1}^N \, {m_j \over r_j} \,,
\label{LMdefn}
\end{equation}
where $r_j \equiv  |\vec y  - \vec y^{(j)} |$.  Regularity (finiteness) of the $Z_I$ and $\mu$ as $r_j \to 0$ requires
\begin{eqnarray}
\ell_j^I  &~=~& -  \coeff{1}{2}\,  C_{IJK} \,
{ k_j^J \, k_j^K  \over q_j} \,,  \qquad j=1,\dots, N \,;\\
m_j &~=~&   \coeff {1}{12}\,C_{IJK} {k_j^I \, k_j^J \, k_j^K \over q_j^2}  ~=~
\coeff{1}{2}\,  {k_j^1 \, k_j^2 \, k_j^3 \over q_j^2} \,,  \qquad j=1,\dots, N \,.
\label{lmchoice}
\end{eqnarray}
In order to obtain solutions that are asymptotic to five-dimensional Minkowski space, $\IR^{4,1}$, one
must take $\varepsilon_0 = 0$, $q_0 =1$  and $k_0^I =0$ in (\ref{KVform}), where $q_0$ is defined in (\ref{q0defn}).  Moreover, $\mu$ must vanish at infinity, and this fixes $m_0$.    As we noted earlier, we also take $Z_I \to 1$ as $r \to \infty$.  

Hence, the solutions that are asymptotic to five-dimensional Minkowski space have:
\begin{equation}
\varepsilon_0 = 0 \,,  \qquad q_0 =1 \,, \qquad  k_0^I =0\,, \qquad   \ell_0^I =1\,, \qquad
 m_0  = -\coeff{1}{2}\, q_0^{-1} \, \sum_{j=1}^N\, \sum_{I=1}^3 k_j^I \,.
\label{fiveDsol}
\end{equation}
The important physical point is that once the $q_j$ and $k^I_j$ have been chosen, the remaining parameters are basically fixed by (\ref{lmchoice}).  The free parameters, $q_j$ and $k^I_j$, determine the magnetic fluxes on the $2$-cycles.

While the functions $Z_I$ and $\mu$ as well as the magnetic fields (\ref{harmtwoform}) are all singular on the critical surfaces, defined by $V=0$, the remarkable thing is that the complete five-dimensional metric (\ref{metform}) and the complete electromagnetic field  (\ref{Aform})  are smooth in a neighbourhood of these critical surfaces.  Indeed from the explicit expressions (\ref{ZIform}) and (\ref{mures}) one can verify that the terms involving negative powers of $V$ cancel out in the complete Maxwell field and metric.

\subsection{Regularity and topology}
\label{RegTop}

The parametrization of the scalar fields in  (\ref{XZrelns}) requires that all the functions, $Z_I$, have the same sign and positive definiteness of the spatial part of the metric  (\ref{metform})  along the $\IR^3$ directions of (\ref{GHmet}) requires that $Z V > 0$.   Thus we must have 
\begin{equation}
Z_I \, V ~>~ 0 \,, \qquad I=1,2,3\,,
\label{ZVpos}
\end{equation}
globally. Constant time slices of the five-dimensional metric gives the four-dimensional metrics
\begin{eqnarray}
{d\hat s}^2_4   &= & - Z^{-2}\, \big( \mu  (d \psi+ A ) + \omega  \big)^2 \nonumber \\
& & ~+~ {Z V^{-1}}\big( d\psi + A \big)^2 + Z V \big(dr^2 +
r^2 d\theta^2 + r^2 \sin^2 \theta \, d\phi^2\big)  \\
&= & {\cQ \over Z^2 V^2} \Big( d\psi + A  - {\mu \, V^2 \over \cQ }\, \omega \Big)^2 ~+~
Z  V \Big( r^2 \sin^2 \theta \, d \phi^2 -{\omega^2  \over \cQ} \Big) 
~+~  Z  V (dr^2 + r^2 d\theta^2) \,,
 \label{dtzero}
\end{eqnarray}
where 
\begin{equation}
Z~\equiv~(Z_1\, Z_2\, Z_3)^{1/3} \,, \qquad \cQ ~\equiv~    Z_1 Z_2 Z_3 V ~-~ \mu^2 \, V^2 \,.
\label{ZQdefn}
\end{equation}
It is also useful to recall that the complete gauge potentials are given by (\ref{Aform}).

\subsubsection{Regularity at critical surfaces}
\label{CritReg}
 
 First observe that at $V=0$, the component pieces of the vector potentials, $Z_I^{-1} (dt +k)$ and $B^{(I)}$, have singular components along $d\psi$.  However, as $V \to 0$, the diverging terms in the complete vector potentials behave as:
\begin{equation}
A^{(I)} ~\sim~ \bigg({ K^I \over V} ~-~ { \mu \over Z_I} \bigg)\, (d \psi +A) 
 ~\sim~  \bigg({ K^I \over V} ~-~ { K^1 \,  K^2 \,K^3
\over  \coeff{1}{2}\,V\, C_{IJK}\,   K^J \,K^K}\bigg) \, (d \psi +A)~\sim~0\,.
\label{AIform}
\end{equation}
Thus $A^{(I)}$ is, in fact, regular on the critical ($V=0$) surfaces.

Similarly, the component parts of the five-dimensional metric, (\ref{metform}), are also singular a $V=0$.    The $(d \psi +A)$ components of the angular momentum vector, $k$,  diverges as $V^{-2}$ while the coefficient of $(dt + k)^2$  is $Z^{-2}$, which vanishes as $V^2$. The cross terms, $ dt \, k$,  remains finite at $V=0$ and the only danger lies in the $(d \psi +A)^2$  terms, and since $Z V$ is finite and positive, the scale of this circle is determined by $\cQ$  in  (\ref{dtzero}) and (\ref{ZQdefn}).  Indeed, it appears from (\ref{ZQdefn}) that this could diverge as $V^{-2}$.  However, there is once again a remarkable cancellation of all the negative powers of $V$  in $\cQ$ and a tedious computation yields the remarkable result:
\begin{eqnarray}
\cQ &~=~&  - M^2\,V^2   - \coeff{1}{3}\,M\,C_{IJK}{K^I}\,{K^J}\,{K^k} - M\,V\,{K^I}\,{L_I}
- \coeff{1}{4} \,(K^I L_I)^2 \nonumber \\
&& \quad +\coeff{1}{6} \, V C^{IJK}L_I L_J L_K +\coeff{1}{4} \,C^{IJK}
C_{IMN}L_J L_K K^M K^N
\label{QasEseven}
\end{eqnarray}
with $C^{IJK} \equiv C_{IJK} = |\epsilon_{IJK}|$.  This is obviously finite on $V=0$ surfaces.  The quantity, $\cQ$,  is, in fact,  the $E_{7(7)}$ quartic invariant written in terms of the eight functions $V, K^I, L_I$ and $2M$.  This quantity is thus duality invariant and for black-hole solutions it determines the horizon area of the four-dimensional black hole obtained by reducing on the $\psi$-fibre.

\subsubsection{Topology, fluxes and  closed time-like curves}

The topology of the five-dimensional metric is determined by that of the four-dimensional base, (\ref{GHmet}). Suppose, for the moment, that $V>0$.  Then representative  cycles, $\Delta_{ij}$, of the non-trivial homology classes can be defined by the $\psi$-fibre over any simple curve between $\vec y^{(i)}$ and $\vec y^{(j)}$.  The fact that $V$ is singular at these points means that the  fibre pinches off thereby defining a compact cycle.    Let $\widehat \Delta_{ij}$ be the cycle $\Delta_{ij}$ with $\vec y^{(i)}$ and $\vec y^{(j)}$ excised.
Since   $\Theta^{(I)}$ is regular  at $\vec y^{(i)}$ and $\vec y^{(j)}$, the flux through $\widehat \Delta_{ij}$ is the same as the flux through $\Delta_{ij}$.   Moreover, the vector potential,  $B^I$ can be globally defined on $\widehat \Delta_{ij}$ because the Dirac strings can be run out through the excised points.   Thus,  the magnetic flux through $\Delta_{ij}$ is  given by:
\begin{eqnarray}
\Pi^{(I)} _{ij} &~\equiv~ & {1 \over 4\, \pi}\,
\int_{\Delta_{ij}} \,
\Theta^{(I)} ~=~ {1 \over 4\, \pi}\, \int_{\widehat \Delta_{ij}} \,
\Theta^{(I)}  ~=~
{1 \over 4\, \pi}\, \int_{\partial \widehat  \Delta_{ij}} \, B^I   \label{fluxnorm} \\ &~=~&
{1 \over 4\, \pi}\,  \int_0^{4\pi} \,  d \psi \, \big( B^I |_{\vec y^{(j)}} ~-~
 B^I |_{\vec y^{(i)}}  \big)  \,.
 \label{basicflux}
 \end{eqnarray}

Now arrange cylindrical polar coordinates, $(\rho,\phi,z)$, so that $\vec y^{(i)}$ and $\vec y^{(j)}$ lie on the $z$-axis at $z=a_i$ and $z=a_j$.  Near the point $\vec y^{(i)}$  one can easily check that the vector potential, $B^I$, in (\ref{Bpot}) is given by 
\begin{equation}
B^I~\sim~ { k^I_i  \over q_i} \, \Big (d \psi  +
q_i \,\Big( {(z -a_i) \over |\vec y  - \vec y^{(i)}| }  +  c_i \Big) \, d \phi \Big)  ~-~
k^I_i \, \Big( { (z -a_i) \over  |\vec y  - \vec y^{(i)}| }  + c_i \Big) \, d \phi   ~\sim~
 {k^I_i  \over q_i} \, d \psi \,.
 \label{Basymp}
\end{equation}
Therefore, the flux is given by 
\begin{equation}
\Pi^{(I)} _{ij} ~=~   \bigg( { k^I_j \over q_j} ~-~ { k^I_i \over q_i} \bigg) \,.
  \label{fluxij}
\end{equation}

This description of the homology cycles and fluxes obviously generalizes to cycles that do not cross critical ($V=0$) surfaces in ambipolar base geometries but the foregoing argument manifestly breaks down if one crosses a critical surface because $\Theta^{(I)}$ is singular.   

However, given the regularity of the complete Maxwell field and five-dimensional metric, one should, of course define the magnetic fluxes by integrating $F^{(I)}$ over the same pieces of the geometry.  Since the full vector potentials are smooth, except at the points $\vec y^{(i)}$ and $\vec y^{(j)}$ one obtains, once again:
\begin{eqnarray}
\widetilde \Pi^{(I)} _{ij} ~\equiv~  {1 \over 4\, \pi}\, \int_{\Delta_{ij}} \, F^{(I)}  ~=~
{1 \over 4\, \pi}\,  \int_0^{4\pi} \,  d \psi \, \big( A^{I} |_{\vec y^{(j)}} ~-~
 A^{I}   |_{\vec y^{(i)}}  \big)  \,.
 \label{fullflux}
 \end{eqnarray}
Using (\ref{Aform}) and (\ref{kansatz}) and recalling that the $\ell_i^I $ and $m_i$ were chosen in (\ref{lmchoice}) so as to make the $Z_I$ and $\mu$ finite at the $\vec y^{(i)}$, one finds
\begin{equation}
\widetilde  \Pi^{(I)} _{ij} ~=~ \Pi^{(I)} _{ij}  ~-~\bigg( { \mu(\vec y^{(j)}) \over Z_I(\vec y^{(j)})} ~-~ {  \mu(\vec y^{(i)}) \over  Z_I(\vec y^{(i)})} \bigg)    \,.
\label{fluxreln}
\end{equation}

The fact that  the $\psi$-fibre pinches off at the $\vec y^{(i)}$  the metric  (\ref{4metric}) means that the $\psi$-circles would become closed time-like curves in the neighborhood of the $\vec y^{(i)}$  because  $ \mu(\vec y^{(i)})$  is generically finite in (\ref{dtzero}).  Thus we must additionally impose the condition 
\begin{equation}
 \mu(\vec y^{(j)}) ~=~ 0\,, \qquad j = 1,\dots, N   \,.
\label{muizero}
\end{equation}
We will discuss this more below, but here we note the important consequence of this for the topology.  

A finite value of  $ \mu(\vec y^{(i)})$  would, of course, have opened up the collapsing curves that define the homology cycles.  Requiring (\ref{muizero}) means that in the full Lorentzian geometry, the $\psi$-circles are pinching off in exactly the manner that was naively suggested by the analysis in the spatial base and thus the naive topology is indeed the exact topology of the full, five-dimensional geometry.   Furthermore, requiring (\ref{muizero}) also means that 
\begin{equation}
\widetilde  \Pi^{(I)} _{ij}  ~\equiv~  {1 \over 4\, \pi}\, \int_{\Delta_{ij}} \, F^{(I)}  ~=~ \Pi^{(I)} _{ij}   ~=~   \bigg( { k^I_j \over q_j} ~-~ { k^I_i \over q_i} \bigg) \,.
\label{fluxisflux}
\end{equation}
Thus the cohomological  fluxes are indeed the naive magnetic fluxes computed on the base and the result remains true even if the cycle crosses a critical surface.

\subsubsection{The bubble equations and closed timeline curves in general}

As we noted above, to avoid closed time-like curves in the neighborhood of the charge centres, one must require (\ref{muizero}).  A rather tedious computation shows that this condition may be rewritten as:
\begin{equation}
 \sum_{{\scriptstyle j=1} \atop {\scriptstyle j \ne i}}^N \,
  {\Gamma_{ij}  \over  |\vec y^{(j)} - \vec y^{(i)} |  } ~=~
-2\, \Big(m_0 \, q_i ~+~  \coeff{1}{2} \sum_{I=1}^3  k^I_i \Big) \,,
\label{BubbleEqns}
\end{equation}
where 
\begin{equation}
\Gamma_{ij} ~\equiv~  q_i \, q_j\, \Pi^{(1)}_{ij} \,   \Pi^{(2)}_{ij} \,  \Pi^{(3)}_{ij} \   \,.
\label{Gammadefn}
\end{equation}
The equations (\ref{BubbleEqns}) are known as the ``bubble equations''     \cite{Bena:2005va,Berglund:2005vb, Bena:2007kg}  or, in the four-dimensional context, ``integrability conditions''  \cite{Denef:2000nb,Bates:2003vx}.  They impose an extra $(N-1)$ conditions\footnote{The sum of the bubble equations is a trivial identity given (\ref{fiveDsol}).} and relate the magnitudes of the magnetic fluxes to the geometric size of  of the configuration.  The moduli space of these solutions is thus $2(N-1)$-dimensional and consists of the relative positions of the  $\vec y^{(i)}$ modulo the $(N-1)$ constraints  (\ref{BubbleEqns}).

Another potential source of CTC's is  the possibility of Dirac-Misner strings in $\omega$. However, from (\ref{omegeqn})  it is relatively easy to see that the absence of Dirac-Misner strings is equivalent to (\ref{muizero}) or  (\ref{BubbleEqns}).

As noted in Section \ref{GlobalStr}, the  complete metric is stably causal if the $t$ coordinate provides a global time function.  In particular,  the condition  (\ref{cond1})  and (\ref{cond2}) now reduce to  \cite{Berglund:2005vb}:
\begin{equation}
- g^{\mu\nu} \partial_{\mu}t \, \partial_{\nu} t = - g^{tt} =  (Z V)^{-1}
(\cQ -  \omega^2) > 0\,,
\label{stabcausal}
\end{equation}
where $\omega$ is squared using the $\IR^3$ metric.

\subsection{Mass, charge and angular  momenta}


As we noted earlier, the spatial metric will be asymptotically flat if the function, $V$,  behaves asymptotically as  $V \sim \frac{1}{r}$. The transformation to spherical polar coordinates at infinity involves setting   $r = {1 \over 4} \rho^2$ and one may then read off the ADM  mass and  the central charges from the coefficients of $\rho^{-2}$ in $g_{00}$ and the electrostatic potentials as described in Section \ref{Normalizations}.  

The electrostatic potentials in (\ref{Aform}) are simply, $-Z_I^{-1}$, and hence the asymptotic charges, $Q_I $, are read off from  the expansion:
\begin{equation}
Z_I ~\sim~ 1 ~+~ \, {Q_I \over 4\, r }~+~ \dots  \,, \qquad  r \to \infty \,.
\label{ZIexpGH}
\end{equation}
These functions have a rather nice expression in terms of the topological fluxes
\begin{equation}
Z_I\,V ~=~ V~-~   \coeff{1}{4} \, C_{IJK} \, \sum_{i, j=1}^N \,
\Pi^{(J)}_{ij} \,   \Pi^{(K)}_{ij} \,    { q_i \, q_j \over  r_i\, r_j} \,,
\label{ZIVexp}
\end{equation}
from which it immediately follows that 
\begin{equation}
Q_I  ~=~ -   C_{IJK} \, \sum_{i, j=1}^N \,
q_i \, q_j \, \Pi^{(J)}_{ij} \,   \Pi^{(K)}_{ij}     \,.
\label{QIanswer}
\end{equation}
This makes the role of the Chern-Simons terms evident in sourcing the electric charge:  the electric charges are quadratics in the topological magnetic fluxes.

Expanding $g_{00}$, one has:
\begin{equation}
-g_{00}~=~  (Z_1 Z_2 Z_3)^{-\frac{2}{3}} ~\sim~ 1 ~-~ \frac{2}{3} \, \sum_{I=1}^3  \, {Q_I \over 4\, r }  \,,
\label{g00expansion}
\end{equation}
and comparing this with (\ref{asympg1})  one finds 
\begin{equation}
M ~=~ \frac{\pi}{4 G_5} \,( Q_1 ~+~ Q_2 ~+~ Q_3) \,. 
\label{MeqlQ}
\end{equation}
It is fairly common to go to a system of units in which the five-dimensional Planck length, $\ell_5$, is unity and this means (see, for example, \cite{Peet:2000hn,Elvang:2004ds}):
\begin{equation}
G_5~=~ \frac{\pi}{4} \,.
\label{NiceUnit}
\end{equation}
In particular, this means that the solution BPS condition takes the simpler standard form:
\begin{equation}
M ~=~ Q_1 ~+~ Q_2 ~+~ Q_3 \,. 
\label{BPSstandard}
\end{equation}

For completeness we also note that the angular momenta can be read off from the asymptotic expansion of the angular momentum vector, $k$:
\begin{equation}
k ~\sim~ {1 \over 4 \,\rho^2} \, \big((J_1+J_2) ~+~ (J_1-J_2) \, \cos \theta   \big) \,
d\psi ~+~ \dots \,,
\label{angmomform}
\end{equation}
where $\theta$ is the polar angle in the flat $\IR^3$ factor in (\ref{GHmet}). One then finds a combinatorial formula for $J_R \equiv J_1 + J_2$:
\begin{equation}
J_R ~\equiv~ J_1 + J_2 ~=~ \coeff{4}{3}\, \, C_{IJK} \, \sum_{j=1}^N q_j^{-2} \,
\tilde  k^I_j \, \tilde  k^J_j \,  \tilde  k^K_j  \,,
\label{Jright}
\end{equation}
where
\begin{equation}
\tilde  k^I_j ~\equiv~ k^I_j ~-~    q_j\, N  \,  k_0^I  \,,
\qquad {\rm and} \qquad k_0^I ~\equiv~{1 \over N} \, \sum_{j=1}^N k_j^I\,.
\label{ktilde}
\end{equation}
Note that as a consequence of taking $q_0 =+1$ in  (\ref{q0defn}) one has
\begin{equation}
\sum_{j=1}^N \,  \tilde  k^I_j  ~=~ 0 \,.
\label{ktildesum}
\end{equation}
We also note that in terms of the parameters $ \tilde  k^I_j $, the expression for the charges takes on a more diagonal form:
\begin{equation}
Q_I ~=~ -2 \, C_{IJK} \, \sum_{j=1}^N \, q_j^{-1} \,
\tilde  k^J_j \, \tilde  k^K_j\,.
\label{QIchg}
\end{equation}

The angular momentum, $J_L \equiv J_1 - J_2$, depends upon the details of  the geometric configuration and may be thought of as a sum of dipole contributions coming from each $2$-cycle:
\begin{equation}
\vec J_L ~\equiv~   J_1 - J_2 ~=~   \sum_{{\scriptstyle i, j=1} \atop {\scriptstyle j \ne i}}^N \,  \vec J_{L\, ij}   \,,
\label{JLnice}
\end{equation}
where 
\begin{equation}
\vec J_{L\, ij} ~\equiv ~ -  \coeff{4}{3}\,q_i \, q_j \, C_{IJK} \,
\Pi^{(I)}_{ij} \,   \Pi^{(J)}_{ij} \,  \Pi^{(K)}_{ij} \, \hat y_{ij} \,,
\label{angmomflux1}
\end{equation}
and $\hat y_{ij}$ are the {\it unit} vectors:
\begin{equation}
\hat y_{ij} ~\equiv ~  {(\vec y^{(i)} - \vec y^{(j)}) \over
 \big|\vec y^{(i)} - \vec y^{(j)}\big| } \,.
 \label{unitvecs}
\end{equation}
If one thinks of 
\begin{equation}
{Q_I}_{ij}   ~\equiv~ -  \coeff{1}{4} \, C_{IJK} \,  q_i \, q_j \, \Pi^{(J)}_{ij} \,   \Pi^{(K)}_{ij}     
\label{QIbits}
\end{equation}
as the contribution to the charges (\ref{QIanswer}) coming from individual bubbles, then \cite{Berglund:2005vb,Balasubramanian:2006gi, Bena:2006kb}
\begin{equation}
\vec J_{L\, ij} ~=~ -  \coeff{16}{3}  \, \sum_{I=1}^3 \,  {Q_I}_{ij} \,  \Pi^{(I)}_{ij} \, \hat y_{ij} \,,
\label{angmomflux2}
\end{equation}
which shows that this part of the angular momentum comes from the ${\bf E} \times {\bf B}$ interactions on each of the bubbles.  

The fact that the individual cycles may be viewed as carrying angular momentum lies at the heart of the semi-classical quantization of the moduli space of solutions \cite{Bena:2007qc,deBoer:2008zn,deBoer:2009un}.

\subsection{Scaling solutions}

While the special properties of scaling solutions is not our main focus here, it would be remiss of us not to give a brief review of the salient features  of this class of solutions since they represent the physically most interesting microstate geometries and in particular they are the ones that closely approximate extremal black holes.

Scaling solutions arise whenever there is a set of points, $\cS$,  for which the bubble equations admit homogeneous solutions  \cite{Denef:2000nb,Bates:2003vx,Bena:2006kb, Denef:2007vg, Bena:2007qc}: 
\begin{equation}
 \sum_{{\scriptstyle j \in \cS } \atop {\scriptstyle j \ne i}} \,   {\Gamma_{ij}  \over  |\vec y^{(j)} - \vec y^{(i)} |  } ~=~  0\,,  \qquad i \in \cS \,.\label{HomBubbleEqns}
\end{equation}
It then follows that such a cluster of points can be scaled: 
\begin{equation}
 \vec y^{(j)} - \vec y^{(i)}  ~\to ~  \lambda  \, ( \vec y^{(j)} - \vec y^{(i)} )    \,, \label{scaling}
\end{equation}
for  $\lambda  \in  \IR$, and one can then examine the limit in which $\lambda \to 0$.  

The geometries are, or course, required to satisfy  (\ref{BubbleEqns}) and not (\ref{HomBubbleEqns}), however, given a solution of  (\ref{HomBubbleEqns}) one can easily make infinitessimal perturbations of the points, $ \vec y^{(i)}$, and if $| \vec y^{(j)} - \vec y^{(i)} |$ is sufficiently small this will generate finite terms on the right-hand side of  (\ref{HomBubbleEqns}) and these can be used to generate solutions to the full bubble equations (\ref{BubbleEqns}).  In this way, the moduli space of physical solutions  that satisfy (\ref{BubbleEqns}) can contain scaling solutions in which a set of points, $\cS$, can approach one another arbitrarily closely. 

The simplest example of this kind of behaviour comes from scaling triangles.  Suppose that $|\Gamma_{ij}|$, $i,j =1,2,3$, satisfy the triangle inequalities:
\begin{equation}
|\Gamma_{13}|   ~<~ |\Gamma_{12}|  ~+~  |\Gamma_{23}|  \qquad { and \ \ cyclic}\,, \label{triangles1}
\end{equation}
which means that we may arrange the points so that 
\begin{equation}
| \vec y^{(j)} -  \vec y^{(i)} |  ~=~  \lambda \, |\Gamma_{ij}|\,, \label{triangles2}
\end{equation}
 for $\lambda \in \IR^+$.  The fluxes  can then usually be arranged so that the homogeneous bubble equations, (\ref{HomBubbleEqns}), are trivially satisfied since they amount to $\pm \lambda^{-1} \mp  \lambda^{-1} =0$.   When the triangle has infinitesimal size, making  infinitesimal deformations of the angles can be used to generate solutions to the original bubble equations  (\ref{BubbleEqns}).  In particular, in a physical solution to (\ref{BubbleEqns}) with three fluxes that obey  (\ref{triangles1}),  one can make the three points approach one another arbitrarily closely by adjusting the angles in the triangle so that they approach the angles in the triangle defined by (\ref{triangles2}).
 
The existence of scaling solutions to the bubble equations, or integrability conditions, was first noted in  \cite{Denef:2000nb,Bates:2003vx}.  However, this seemed to be a rather singular limit but it was subsequently shown that, from the perspective of five-dimensional supergravity, this limit is not only  non-singular but also defines perhaps the most important class of physical solutions \cite{Bena:2006kb, Denef:2007vg, Bena:2007qc}.   

Suppose that we have a scaling cluster, $\cS$, that is centred on the origin, $r =0$.  Let $\epsilon$ be the largest separation (in $\IR^3$) between points in $\cS$ and let $\eta$ be the smallest distance from a point in $\cS$ and a point, $ \vec y^{(i)}$,  that not in $\cS$.  Assume that $\epsilon <\! <\! < \eta$ and,  for simplicity,  suppose that the total geometric charge of the cluster is unity:  $q_\cS  \equiv \sum_{i \in \cS} q_i = 1$.   In the intermediate range of $r$ in which, $\epsilon \ll r \ll \eta$, one has $V \sim \frac{1}{r}$ and all the other functions $K_I$ and $L_I$ behave as $\cO(r^{-1})$.   This means that, in the intermediate region, $Z_I \sim \frac{Q_{I , \cS}}{4 r}$. where  the $Q_{I, \cS}$ are the electric charges associated with the scaling cluster.  Using this in (\ref{metform}) and (\ref{GHmet}) we see that the metric in the intermediate region becomes:
\begin{equation}
ds_5^2 ~=~  - \frac{16\,r^2}{a^4}  (dt +k)^2 ~+~      \frac{a^2}{4} \, \frac{dr^2}{r^2}   ~+~  \frac{a^2}{4} \, \big[ ( d \psi +  \cos \theta d \phi ) ^2 +  d\theta^2 + \sin^2 \theta d \phi^2  \big]   \,,
\label{Intermet}
\end{equation}
where $a =(Q_{1, \cS}Q_{2, \cS}Q_{3, \cS})^{1/6}$.   This is the metric of an $AdS_2 \times S^3$ throat of a rotating, extremal black hole.

There are several important consequences of this result.  First,  such scaling  clusters look almost exactly like extremal black holes  except that they ``cap off'' in a collection of bubbles just above\footnote{From the perspective of an infalling observer.} where the horizon would be for the extremal black hole.  Moreover, while it appears, from the perspective of the $\IR^3$ base, that the bubbles are collapsing in the scaling limit, they are, in fact, simply creating an $AdS$ throat and descending down it as it forms.  The physical size of the bubbles  approaches a large, finite value whose scale is set by the radius, $a$, of the $S^3$  of the throat, which corresponds to the horizon of the would-be black hole.  Thus the scaling microstate geometries represent deep bound states of bubbles that realize the goal of creating a smooth, solitonic solutions that look  like BPS black holes.     One obtains similar results for black rings from scaling clusters  whose net geometric charge, $q_\cS$, is zero.

The fact that one can adjust classical parameters so that the scaling points approach one another arbitrarily closely means that the $AdS$ throat can be made arbitrarily deep.  However, the angular momentum
(\ref{JLnice}) depends, via (\ref{angmomflux1}), upon the details of locations of the points and  when angular momentum is  quantized this will lead to a discretization of the moduli space and will limit the depth of simple scaling solutions like those based on scaling triangles  \cite{Bena:2007qc}.  More generally, it was proposed in  \cite{Bena:2007qc} and then proven in \cite{deBoer:2008zn} that the individual contributions, $\vec J_{L\, ij}$ in (\ref{angmomflux1}) must be separately quantized and so, upon quantization, the classical moduli space is completely discrete.  This has the very interesting physical consequence that even though very long, deep throats are macroscopic regions of space time in which the curvature length scale can be uniformly bounded to well above the Planck scale, quantum effects can wipe out such regions of space-time.

\subsection{Some simple examples}

\subsubsection{Two centres:  $AdS_3 \times S^2$}
\label{AdSS}
  
Our first example \cite{Denef:2007yt,deBoer:2008fk,Bena:2010gg} has  equal and opposite geometric charges and $\varepsilon_0 =0$ in (\ref{KVform}).  We now show that the metric is that of $AdS_3 \times S^2$.  More generally, if the total geometric charge, $q_0$ in (\ref{q0defn}), vanishes then the metric will be asymptotic to global $AdS_3 \times S^2$.   In particular, we make contact with the discussion  at the end of Section 3.

We locate the two centres on the $z$-axis at $ z= \pm a$ and define:
\begin{equation}
r_\pm ~\equiv~   \sqrt{\rho^2 ~+~ (z\mp a)^2 } \,,
\end{equation}
where $(z, \rho, \phi)$  are cylindrical polar coordinates on the $\IR^3$ base.
Take the harmonic functions to be 
\begin{eqnarray}
V &=&   \Big({ 1 \over r_+} ~-~ {1 \over r_-} \Big)\,, \qquad  K ~=~
k\, \Big({ 1 \over r_+} ~+~ {1 \over r_-} \Big)  \,,  \\
\qquad  L&=&-k^2 \, \Big({ 1 \over r_+} ~-~ {1 \over r_-} \Big) \,, \qquad  M ~=~
- {2\, k^3 \over a}~+~ \frac{1}{2} \,k^3  \, \Big({ 1 \over r_+} ~+~
{1 \over r_-} \Big)\,,
\end{eqnarray}
where the constant in $M$ has been chosen so as to make the metric regular at infinity.

The vector potentials for this solution are then:
\begin{equation}
A ~=~   \Big({(z -a) \over r_+} - {(z +a) \over r_-} \Big) \, d\phi  \,, \qquad \omega ~=~ -{2\, k^3 \over a} \,
{\rho^2 + (z-a +r_+)(z+a - r_-)  \over r_+ \, r_-}  \, d\phi  \,.
\end{equation}
The five-dimensional metric is then:
\begin{equation}
ds_5^2 ~\equiv~ - Z^{-2} \big(dt+ \mu (d\psi+A) + \omega\big)^2 ~+~
Z\, \big(   V^{-1} (d\psi+A)^2 ~+~ V(d\rho^2 + \rho^2 d\phi^2 + dz^2)  \big)\,,
\label{fivemetric}
\end{equation}
where
\begin{eqnarray}
Z &=&  V^{-1} K^2 + L ~=~ -  { 4\, k^2    \over (r_+ - r_- )}\,,  \\
\mu &=&  V^{-2} K^3 +  \coeff{3}{2}\, V^{-1} K \, L + M ~=~ {4\, k^3}  \,
{  (r_+ + r_- ) \over (r_+ - r_- )^2} ~-~  {2\, k^3 \over a} \,. \nonumber
\end{eqnarray}

To map this onto a more standard form of $AdS_3 \times S^2$ one must make a
transformation to oblate spheroidal coordinates like those employed in \cite{Prasad:1979kg}
to map positive-definite two-centred space onto the Eguchi-Hanson form:
\begin{equation}
 z =  a\, \cosh 2\xi \,\cos \theta \,, \qquad  \rho =  a\, \sinh 2\xi \, \sin \theta \,, \qquad
 \xi \ge 0\,, \ \ 0 \le \theta \le \pi \,.
  \label{coordsa}
 \end{equation}
In particular, one has $r_\pm =  a (\cosh 2\xi  \mp \cos \theta)$.  One then rescales and shifts the remaining variables according to:
\begin{equation}
 \tau ~\equiv~   \coeff{a}{8\, k^3}\,  t \,, \qquad \varphi_1 ~\equiv~   \coeff{1}{2} \, \psi -
 \coeff{a}{8\, k^3}\,  t   \,, \qquad \varphi_2 ~\equiv~ \phi -   \coeff{1}{2} \, \psi +
 \coeff{a}{4\, k^3}\,  t   \,,
 \label{coordsb}
 \end{equation}
and the five-dimensional metric takes the standard $AdS_3 \times S^2$ form:
\begin{equation}
ds_5^2 ~\equiv~ R_1^2 \big[ - \cosh^2\xi \,  d\tau^2 + d\xi^2 +  \sinh^2 \xi \, d\varphi_1^2 \big] ~+~  R_2^2 \big[   d \theta ^2 + \sin^2\theta  \, d\varphi_2^2 \big]  \,,
 \label{AdS3S2}
 \end{equation}
with
\begin{equation}
  R_1~=~  2 R_2 ~=~ 4 k \,.
 \label{Radii}
 \end{equation}
Note that the first factor in the metric is {\it global} $AdS_3$ with $-\infty < \tau < \infty$.

Note that
\begin{equation}
 g_{\psi \psi} ~=~  \coeff{1}{4}\, R_1^2 \,  \sinh^2\xi ~+~ \coeff{1}{4}\, R_2^2 \,  \sin^2\theta  \,.
 \label{gpsipsi}
 \end{equation}
Thus the Killing field, ${\partial \over \partial \psi}$ has two fixed points, or nuts, at $\xi = 0, \theta =0$ and  $\xi = 0, \theta =\pi$, that is, at $r_\pm =0$.  It follows that the base manifold, $\cB$, has Euler characteristic $2$, consistent with its topology being $\IR^2 \times S^2$.

Also note that the time-like Killing vector, $\bK = \frac{\partial}{\partial t}$, of the original space-time metric  (\ref{metform}) is related, in this example, to the Killing vectors of $AdS_3 \times S^2$ by (\ref{ads3s2}). In this sense, the only unusual feature of the critical surface, $V=0$, is the behaviour of the family of ``observers'' with worldlines defined by the integral curves of  $\bK = \frac{\partial}{\partial t}$.  More generally, critical surfaces  occur between between pairs of geometric charges of opposite sign and the smoothness of the solution across  critical surfaces will follow for the same basic reason that we have described here. 

\subsubsection{Three centres}

We now consider an example in which $\varepsilon_0 =0$ again but the total geometric charge, $q_0$ in (\ref{q0defn}), is $1$.    This means that the space-time is asymptotically Euclidean.

Take
\begin{equation}
 V = {1  \over r_+} ~-~ {1  \over r_0}~+~  {1  \over r_-}  \,,
\label{simpVform}
\end{equation}
and
\begin{equation}
K^I ~=~ K  = {k   \over r_+} ~+~ {k  \over r_0}~+~  {k   \over r_-}  \,,
\label{simpKform}
\end{equation}
where 
\begin{equation}
r_\pm  ~=~ \sqrt{\rho^2 ~+~ (z \mp 1)^2}  \,, \qquad r_\pm  ~=~ \sqrt{\rho^2 ~+~ z^2}  \,.
\label{rdefn}
\end{equation}

The remaining functions are then
\begin{equation}
L_I~=~ L  ~=~1 ~-~  {k^2   \over r_+} ~+~ {k^2  \over r_0}~-~  {k^2  \over r_-} ~=~ 1 ~-~  k^2\, V  \,,
\label{simpLform}
\end{equation}
\begin{equation}
M  ~=~  -\frac{9}{2}\, k ~+~  \frac{1}{2}\, \Big( {k^3  \over r_+} ~+~ {k^3    \over r_0}~+~  {k^3    \over r_-} \Big)  ~=~ -\frac{9}{2}\, k ~+~   \frac{1}{2}\, k^2 \, K \,,
\label{simpMform}
\end{equation}

Normally (\ref{BubbleEqns}) gives complicated rational equations relating flux parameters to positions but for this problem it collapses to:
\begin{equation}
k  ~=~ \frac{\sqrt{3}}{2}  \,.
\label{Bsoln}
\end{equation}

One then has
\begin{align}
Z\, V  ~=~ & K^2 ~+~ L\, V  ~=~  V ~+~ {4\, k^2 \over r_0} \Big(  {1 \over r_-}~+~   {1 \over r_+}  \Big)\\ 
~=~  & {1  \over r_+} ~-~ {1  \over r_0}~+~  {1  \over r_-} ~+~ {3\over r_0} \Big(  {1 \over r_-}~+~   {1 \over r_+}  \Big) \\ 
~=~  & {r_0 -   r_+ + 1\over r_0 \, r_+} ~+~ {1  \over r_-}~+~  {3  \over r_- \, r_0 } ~+~   {2  \over r_0  \, r_+ }
\label{ZVres}
\end{align}
The first term is positive definite because of the triangle inequality and so one has $ Z V >0$ globally.

Set 
\begin{equation}
  V_0 ~\equiv~  {1  \over r_0} \,, \qquad  V_1 ~\equiv~  {1  \over r_+}~+~  {1  \over r_-}  \,,
\label{V01defns}
\end{equation}
then $\mu$ in (\ref{kansatz}) can be simplified to
\begin{equation}
\mu ~=~ V^{-2} \,\big( 4 k^3 \, V_0  V_1 (V_0 + V_1)  ~+~ 3 k \,   (V_1 - V_0) (2\,V_0-  V_1)    \big)   \,, 
\label{musimp}
\end{equation}
and $\cQ$ may be written
\begin{equation}
\cQ ~=~- \frac{27}{4}  \,  (V_0  V_1 - V_1 - 2  V_0 )^2  ~+~ 63 \, V_0 V_1 +  (V_1 - V_0)    \,.
\label{Qsimp}
\end{equation}

Finally, $\omega$  in (\ref{kansatz}) can be written:
\begin{equation}
\omega =  - \frac{4 k^3 }{r_0}\, \Big[ \frac{1}{r_+} \big(  \rho^2 + (z-r_0)(z-1 + r_+) \big)  - \frac{1}{r_-} \big(  \rho^2 + (z+r_0)(z + 1 - r_-) \big)     \Big] \, d \phi     \,.
\label{omres1}
\end{equation}
Observe that $\omega$ vanishes identically on the $z$-axis ($\rho =0$) as it must if one is to avoid CTC's  (indeed it vanishes quadratically in  $\rho$).

\begin{figure}[t]
 \centering
    \includegraphics[width=10.5cm]{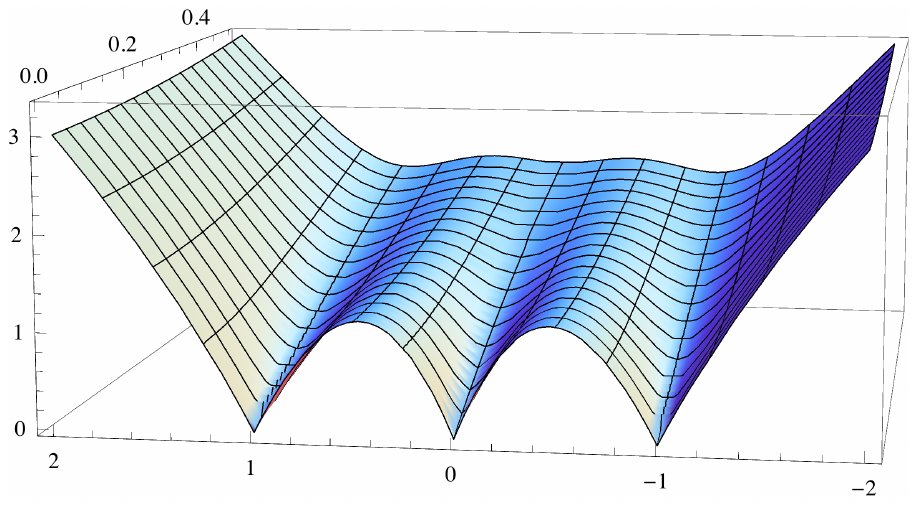}
    \caption{\it \small Graph of ${\cQ \over Z^2 V^2}$ for $-2 \le z \le 2$, $0 \le \rho \le 0.5$. Observe that it is globally non-negative.}
\label{fig1}
\end{figure}

If one plots  ${\cQ \over Z^2 V^2}$, as shown in Fig. \ref{fig1}, one easily  see that it  is non-negative and vanishes only when  $\rho=0$ and either $z=0$ or $z=\pm 1$.  Thus (\ref{Qsimp}) has the proper behaviour.

\begin{figure}[t]
 \centering
    \includegraphics[width=10.5cm]{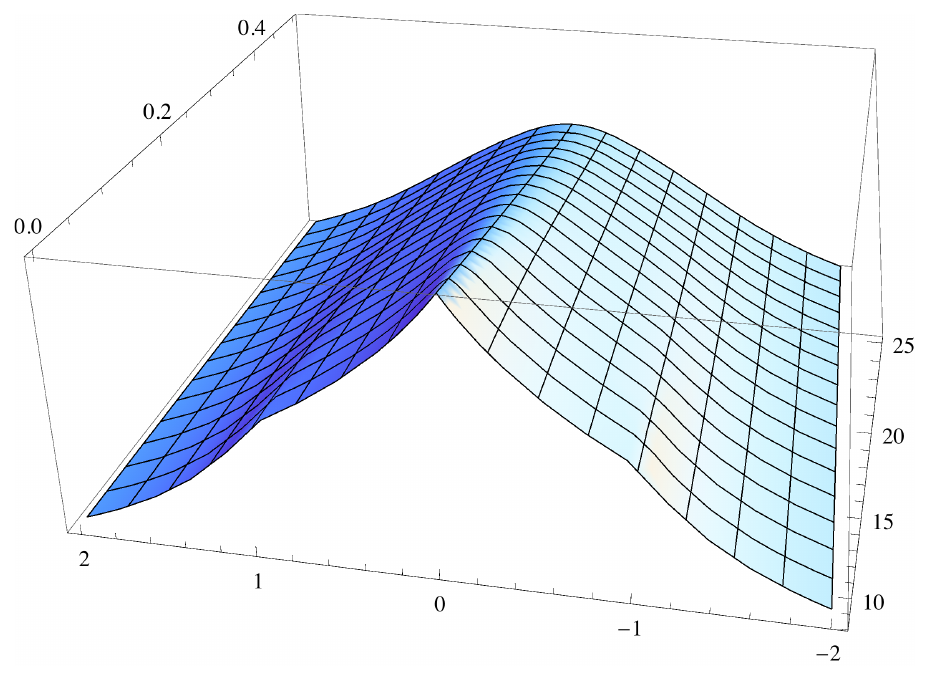}
    \caption{\it \small Graph of $-g^{tt}$ given in (\ref{stabcausal}) for $-2 \le z \le 2$, $0 \le \rho \le 0.5$. Observe that $-g^{tt}>0$ everywhere and so the space-time is stably causal. The cusps at $\rho=0$, $z=\pm1$ and $\rho=0$, $z=0$ are a coordinate artifacts arising from the fact that $\rho$ and $z$ are not good coordinates at these points. }
\label{fig2}
\end{figure}

If one plots  $(Z V)^{-1} (\cQ -  \omega^2)$, as shown in Fig. \ref{fig2},  one may verify that it is strictly positive and regular and so (\ref{stabcausal}) is satisfied. One finds that  $g^{tt}$ has discontinuous first derivatives in $z$ at $z=0$ and $z=\pm 1$ for $\rho=0$.  These are coordinate artifacts  inherited from the fact that $\rho$ and $z$ are not good coordinates in the neighborhood of $\rho=0$ and $z=0, \pm 1$. These apparent kinks are removed by the usual $r_j  \to \frac{1}{4} R_j^2$ change of variable.  

\goodbreak

\goodbreak

\section{A non-BPS Soliton that violates the BPS bound } 

There are only rather few  known examples of non-BPS, non-extremal solitons \cite{Jejjala:2005yu,Bena:2009qv,Bobev:2009kn,Dall'Agata:2010dy,Bossard:2011kz,Bossard:2012ge} and most of these involve a single topological bubble.  The difficulty is, of course,  largely because one has to solve equations of motion rather than the much simpler BPS equations.  Here we will  re-analyse a hitherto puzzling ``soliton'' solution of five-dimensional minimal supergravity \cite{Compere:2009iy}  using the methods and ideas of the present paper.  The approach to finding this soliton is rather reminiscent of the construction of the soliton in \cite{Jejjala:2005yu,Bena:2009qv,Bobev:2009kn} except that the solution of  \cite{Jejjala:2005yu} is only regular in six dimensions while the solution of   \cite{Bena:2009qv,Bobev:2009kn,Compere:2009iy} are smooth in five-dimensions.  A striking feature of the solutions of  \cite{Compere:2009iy} is that they violate the BPS bound for the mass in terms of the electric charge  (a proof of which for minimal supergravity using spinors was  given in \cite{Gibbons:1993xt}).  We shall argue that this anomalous behaviour arises because this solitonic spacetime does not admit a spin structure.
   
\subsection{The solution}

The  Lagrangian of five-dimensional minimal supergravity is obtained from   (\ref{5daction}) by setting all the vector fields equal\footnote{With these choices of normalization our five-dimensional action matches precisely the ungauged action in \cite{Compere:2009iy}.}, $A^I = \frac{1}{\sqrt 3} A$, and trivializing the scalars by setting $X^I =1$.    

The starting point for constructing the soliton is the black hole solution of  \cite{Chong:2006zx} in gauged minimal supergravity, which we will  immediately specialize  to the ungauged theory by setting the gauge coupling, $g$, to zero.  If $g=0$  then it is claimed in \cite{Compere:2009iy}  that regularity requires that one set the parameters $a$ and $b$ of  \cite{Chong:2006zx}  to be of equal magnitude and the parameter $q$ must be negative.  We choose $a=b$ and the metric then becomes:
\begin{align}
ds ^2 ~=~  &   -\Big(  1 - \frac{2m}{\rho^2} +  \frac{q^2}{\rho^4}  \Big) ~-~  \frac{2}{\rho^2}  \Big(  (q+2m)  -  \frac{q^2}{\rho^2}  \Big)  \,dt \, \omega   ~+~  \frac{1}{\rho^2} \, \Big(  2(q+m)  -  \frac{q^2}{\rho^2}  \Big) \, \omega^2 \nonumber  \\ 
&~+~  \frac{\rho^2 r^2 dr^2}{W} ~+~ \rho^2  \big( d\theta^2 +  \sin^2 \theta d \phi_1^2 +   \cos^2 \theta d \phi_2^2 \big) \,,\label{nBPSmet}
\end{align}
where
\begin{equation}
\rho^2 ~\equiv~  r^2 + a^2 \,,\qquad \omega  ~\equiv~  a(\sin^2 \theta d \phi_1 +   \cos^2 \theta d \phi_2)\,,  \qquad W  ~\equiv~  \rho^4  +  q^2 +2 a^2 q  - 2 m r^2 \,, 
\label{rhoWomdefns}
\end{equation}

The soliton is found by continuing to negative values of $r^2$ and so we define  $X=\rho^2  = r^2 +a^2$ so that $2rdr = dX$. The spatial metric (obtained by setting  $t=0$ ) is, in fact, of cohomogenity one and  admits an action of
$U(2)$ whose principal   orbits $X= {\rm constant}$  are squashed 3-spheres.  
To make this manfest we  define  $\phi_1 =(\psi + \phi)/2 $ and $\phi_2 = (\psi -\phi)/2$ and introduce the standard left-invariant one-forms noting that   $\omega =  \frac{a}{2} \sigma_3$.  The metric is therefore given by:  
\begin{align}
ds ^2 ~=~ & 
 \frac{X dX ^2 }{4\big (X^2 -2mX + 2 a^2 (q+m) + q^2  \big ) } ~+~ \frac{X}{4} \big(\sigma _1^2 +\sigma _2^2 +  \sigma _3^2 \big) \nonumber \\ 
& ~-~ \Big( 1- \frac{2m}{X} + \frac{q^2}{X^2 }  \Big ) \, dt ^2~-~ 2 \Big( \frac{2m+q }{X} - \frac{q^2 }{X^2}  \Big ) \, a \,\sigma_3\, dt   ~+~ \frac{a^2}{4} \Big( 2\frac{m+q}{X} - \frac{q^2 }{X^2}   \Big )\, \sigma_3 ^2 \,.
\end{align}
The Maxwell field is relatively simple:
\begin{equation}
A ~=~ q\,  \frac{\sqrt{3}}{X} \, \Big( dt - \frac{a}{2} \sigma_3\Big) \,. \label{Apot1}
\end{equation}

We  define
\begin{equation}
Q ~\equiv~   -  \frac{q}{a^2} 
\end{equation}
and note that, according to \cite{Compere:2009iy}, regularity requires that $Q>0$.
In \cite{Compere:2009iy} it is also pointed out that if $m$ is chosen so that
\begin{equation}
 m ~=~ -\half \, q \big (1+\sqrt{Q}\big) ~=~ \half \, a^2 Q\,  \big(1+\sqrt{Q}\big) \,, 
\end{equation}
with 
\begin{equation}
0< -a^2 q < a^4 \quad \Leftrightarrow \quad 0 <Q<1\,, 
\end{equation}
then one can arrange that   
\begin{equation}
g_{\psi t}(X_0) ~=~ 0\,,\qquad  g^{XX}( X_0) ~=~ 0\,,
\qquad g_{\psi \psi}( X_0) ~=~ 0  
\end{equation}
at 
\begin{equation}
X~=~ X_0~\equiv~ \sqrt{-a^2 q} ~=~ a^2 \sqrt{Q}    \,.
\end{equation}
Note that for this to happen, $q$ must be negative and that $m$ and $2m+q$  are both  positive
but $m+q$ is negative. The position, $X_0$, is also positive but at $X=X_0$, $r^2$ is negative.

In the obvious notation we indeed find the following
\begin{align}
g_{tt}~=~ & - \Big( 1- \frac{2m}{X} + \frac{q^2}{X^2 }  \Big ) ~=~ - X^{-2} \big(X^2 - a^2 Q\big(1+\sqrt{Q}\big)X + a^4 Q^2\big) \,,  \label{gttform} \\
g_{\psi t}~=~ &  a - \Big( \frac{2m+q }{X} - \frac{q^2 }{X^2}  \Big ) ~=~ - X^{-2} \big(a \sqrt{Q}\big)^3 \big(X- a^2 \sqrt{Q}\big) \,, \\
g_{\psi \psi} ~=~ &  \frac{X}{4} + \frac{a^2}{4} \Big( 2\frac{m+q}{X} - \frac{q^2 }{X^2}   \Big ) ~=~ - X^{-2}\big(X- a^2 \sqrt{Q}\big) \big(X^2 +  a^2 \sqrt{Q} X +   a^4 Q \sqrt{Q}\big)\,, \\
g^{XX}~=~ & 4\,X^{-1} \big (X^2 -2mX + 2 a^2 (q+m) +q^2 \big )\\  
& \qquad \qquad\qquad\qquad\qquad \ \ \, ~=~ 4\,X^{-1} \big(X- a^2 \sqrt{Q}\big) \big(X +   a^2  \sqrt{Q}\big(1 -\sqrt{Q} -Q \big)\big)  \label{gXXup}
\end{align}
and 
\begin{equation}
g^{tt} ~=~  \frac{g_{\psi \psi }}{g_{tt}g_{\psi \psi} - g_{\psi t}^2 }   ~=~ \frac{\big(X^2 +  a^2 \sqrt{Q} X +   a^4 Q \sqrt{Q}\big) } { X \bigl (X +   a^2  \sqrt{Q}\big(1 -\sqrt{Q} -Q \big)  \bigr )} \label{gttup}  \,.
\end{equation}

The fact that $g_{\psi t}(X_0)  =  g_{\psi \psi}( X_0) = 0$ means that  the $\psi$-fibre is pinching off at $X= X_0$ leaving a finite sized $S^2$ ``bolt,'' or ``bubble.''   Smoothness requires that we make sure that the metric has  no conical singularities at the bolt and so we need to examine the metric in the neighborhood of $X= X_0$.

\subsection{Regularity and properties of the solution}

\subsubsection{Removal of conical singularities}

Setting $X=X_0+x $ and expanding $x$, the spatial metric transverse to the bolt has the form 
\begin{equation}
ds_2^2 ~=~  u\,  \frac{dx ^2}{x}  ~+~  v \, x \, d \psi ^2 ,
\end{equation} 
where 
\begin{align}
u   ~=~&  {g^\prime_ {XX}  } (X_0)  ~=~ \frac{1}{8}  \Big ( 1- \frac{m}{X_0}\Big )^{-1} ~=~ 
\frac{1}{4 (2- \sqrt{Q} - Q )} \,,  \\
 v ~=~ & g_{\psi \psi} ^\prime (X_0)  ~=~ \frac{1}{4} \Big(  1-2 a^2 \frac{m+q}{X_0^2}      + 2a^2 \frac{q^2}{X_0^3}             \Big)  ~=~  \frac{1}{4} (2+ \sqrt{Q} )   \,, 
\end{align}
and $\,^\prime$ denotes differentiation with respect to $X$.

If we define a radial variable 
\begin{equation}
y= 2\sqrt{u x} 
\end{equation}  
we find
\begin{equation}
ds^2 _2 = d y^2 + y^2 \frac{v}{4u} d \psi^2
\end{equation}

and therefore in order to eliminate a conical singularity
we require
\begin{equation}
\psi \in (0, 4\pi \sqrt{ \frac {u}{v} }     ]
\end{equation}
On the other hand asymptotic flatness requires
\begin{equation}
\psi \in ( 0, 4 \pi] 
\end{equation}
and thus consistency requires
\begin{equation}
\frac {v}{u} ~=~ (2+ \sqrt{Q} )(2- \sqrt{Q} - Q )  ~=~  1\,.
\end{equation}
This cubic has a unique solution in the range $0<Q<1$   and it is at  $Q \approx 0.7733184...$ \,.

\subsubsection{Ergo-regions and stable causality}

From  (\ref{gttform})  one can write 
\begin{equation}
- g_{tt} ~=~ X^{-2} \big(( X - a^2 Q)^2 +Q  (1- \sqrt{Q})X \big) 
\end{equation}
which is manifestly positive for $0 < Q <1$ and hence there are no ergo-regions.  

Moreover, from  (\ref{gttup}) one can easily see that $g^{tt} >0$ for  $X > X_0 = a^2 \sqrt{Q}$, with $0 < Q <1$ and hence the space-time is stably causal.

\subsubsection{The Field strength}

The field strength $F=dA$ with $A$ given by (\ref{Apot1}) is simply  
\begin{equation}
F ~=~ - q\,\sqrt{3}\,  X^{-2} \, dX \wedge \Big( dt - \frac{a}{2} \sigma_3\Big) ~-~  2 \sqrt{3} \,a q\, X^{-1} \, \sigma_1 \wedge \sigma_2 \label{Fform1} \,.
\end{equation}
Since the $2$-sphere always has finite size for $X \ge X_0$, the second term is globally regular and gives a finite contribution to  
\begin{equation}
F^2 ~\equiv~ F_{\mu \nu} F^{\mu \nu}  \,.
\end{equation}

The first term in (\ref{Fform1}) is similarly smooth, but it might give a divergent contribution to $F^2$ because the fibre is pinching off.  One can easily check that $g^{\psi \psi}$ diverges as $(X- X_0)^{-1}$ at $X=X_0$, however from (\ref{gXXup}) one sees that $g^{X X}$ vanishes as $(X- X_0)$ and so the contribution to $F^2$ is also finite.
Thus $F$ is well behaved.

 \subsubsection{Mass and Charge: Violation of the BPS bound}

According to \cite{Compere:2009iy} the mass $M$ and electric charge $Q_E$ are given by
\begin{equation}
M= \frac{3 \pi m}{4G} \,,\qquad Q_E = - \frac{\sqrt{3} \pi a^2 Q}{4G}
\end{equation} 
Thus
\begin{equation}
\frac{\sqrt{3} Q_E}{M} =- \frac{2}{1+\sqrt{Q}} 
\end{equation}
which is in contradiction with the usual BPS bound 
 \cite{Gibbons:1993xt}
\begin{equation}
M \ge \sqrt{3}|Q_E| \,.
\end{equation}
This seems very mysterious but the explanation 
 lies in the topology and the fact that the space-time does not have a spin structure, as we will now demonstrate.

Since  the space-time is stably causal,  the local section, or base ${\cal B}$, given by  $t={\rm constant} $ is a Cauchy surface on which the induced  metric: 
\begin{equation}
ds^2 _4=  \frac{X dX ^2 }{4\bigl (X^2 -2mX + 2 a^2 (q+m) + q^2  \bigr ) } +     \frac{X}{4} \bigl(\sigma _1^2 +\sigma _2^2) +  \frac{1}{4}\Big(X + a^2 \Big( 2\frac{m+q}{X} - \frac{q^2 }{X^2}   \Big )\Big ) 
\sigma_3^2
\end{equation}
is positive definite.
 
Since there is an action of $U(2)$ with a two-dimensional fixed point set, we can compare with  the $U(2)$ invariant form of the Fubini-Study metric on ${\Bbb C} {\Bbb P} ^2$ given in \cite{Gibbons:1978zy}
\begin{equation}
ds ^2 =  \frac{dr ^2}{1+ \frac{\Lambda r^2}{6} } 
+ \frac{r^2} {4 (1+ \frac{\Lambda r^2}{6} )^2 } \sigma _3 ^2 +
 \frac{r^2}{4 (1+ \frac{\Lambda r^2}{6} )} \bigl( \sigma _1^2 + \sigma _2 ^2   \bigr )    \,,
\end{equation}
where $\Lambda$ is the cosmological constant. If we set $r=\frac{1}{X-X_0}$ then the origin, $r=0$,  of ${\Bbb C} {\Bbb P}^2$ maps to  infinity $X=\infty$ and the bolt, or complex line,  at $r=\infty$ maps to  the bolt at $ X=X_0$.
Thus the base, $\cal B$,  has the topology, but of course not the metric,  of   ${\Bbb C} {\Bbb P} ^2 \setminus {0} $. The bolt,  or bubble,  corresponds to a complex line $ {\Bbb C} {\Bbb P} ^1$ in ${\Bbb C} {\Bbb P} ^2 $. Notoriously every line in the complex projective plane intersects every other line once and only once and therefore 
the self-intersection number of the bolt is unity, which is an odd number, and so the  second Stiefel-Whitney class  of  $\cal B$ is non-vanishing and hence the spacetime does  not admit a spin structure. 

This presumably accounts for the breakdown of the spinorial proof given in \cite{Gibbons:1993xt} of the 
BPS bound.

\section{Topology, Fluxes, Charge and the Smarr Formula for BPS Solitons} 

\subsection{The Smarr formula revisited}
\label{Smarr2}

We now apply the results of Section \ref{Smarr1} to the general class of examples that we have just described.  
First, it follows immediately from (\ref{Aform}) and (\ref{KdotF}) that 
\begin{equation}
\lambda^I  ~=~    Z_{I}^{-1} ~-~ \alpha^I  \,,
\label{lambdaform}
\end{equation}
where $ \alpha^I$ are constants. Indeed, with the boundary conditions  (\ref{Zasymp}), one should take the $\alpha^I = 1$.  It is only with this choice that the boundary terms at infinity involving $\lambda^I$ can be dropped in (\ref{Mform2})  to arrive at the result (\ref{ADMmassres}) and ultimately at   (\ref{Manswer}).  In spite of this, we will retain $ \alpha^I$ because they will prove useful in elucidating the computation.

One can then show that
\begin{align}
i_K G_I ~+~  \coeff{1}{2}  \,C_{IJK} \,   \lambda^J \, F^{K}  ~=~ -  \coeff{1}{2}  \, d \big( Z_I Z^{-3} \, (dt + k) \big)  ~-~  \coeff{1}{2}  \,C_{IJK} \,  \alpha^J \, F^{K}   \,, 
 \label{simpKdotG}  
\end{align}
where one should recall the definition of $Z$ in (\ref{XZrelns}).  Hence one has
\begin{align}
\Lambda_I  ~=~ -  \coeff{1}{2}  \, d \big( Z_I Z^{-3} \, (dt + k) \big) \,, \qquad H_I ~=~  - \coeff{1}{2}  \,C_{IJK} \,  \alpha^J \, F^{K}   \,.
 \label{Lambdadefn}  
\end{align}

We must show that the vector field, $\Lambda_I $, is globally well-defined and smooth.  There are several potential sources of singular behaviour: (i) if one of the $Z_J$ vanishes, (ii) when $V=0$ and (iii) when $r_j =0$.  First we note that $Z_I V >0$ and $Z_I$ is finite when  $r_j =0$ and thus $Z_I$ never vanishes. Since $Z_I$ is finite when  $r_j =0$, and the bubble equations require that $k$ vanishes when   $r_j =0$, there are no singularities at $r_j =0$.  Finally, near $V=0$, the vector $k$ diverges as $V^{-2}$ but $Z_I Z^{-3} \sim \cO(V^2)$ and so $\Lambda_I$ is smooth in a neighborhood of these surfaces.

The mass formula thus reduces to:
\begin{align}
  M ~=~&      \frac{1}{32 \pi G_5} \,   \int _{ \Sigma}   \Big[  Q^{IJ}\, H_{I \rho \sigma} \, {G_J}^{\rho \sigma \nu }\Big]  \, d \Sigma_{\nu} 
~=~      - \frac{1}{64 \pi G_5} \,   C_{ILM} \,  \alpha^L \,  \int _{ \Sigma}   \Big[  Q^{IJ}\, F^M_{ \rho \sigma} \, {G_J}^{\rho \sigma \nu }\Big]  \, d \Sigma_{\nu}    \nonumber \\  
~=~&      -  \frac{1}{128 \pi G_5} \,   C_{IJK} \,  \alpha^I \,  \int _{ \Sigma}    \epsilon_{\mu \alpha \beta \rho \sigma} F^{J \alpha \beta } \, F^{K\rho \sigma}  \, d \Sigma^{\mu} 
\label{simpM1} \\
~=~&      -  \frac{1}{8 \pi G_5} \,  \int _{ \Sigma}  \, \nabla_\rho \, \Big[ \alpha^I  Q_{IJ} \, {F^{J \rho}}_\mu \, \Big]   \, d \Sigma^{\mu}  
~=~    -  \frac{1}{8 \pi G_5} \,   \int _{\infty}  \, \alpha^I  \,Q_{IJ} \, {F^J}_{\mu \nu} \,   \, d \Sigma^{\mu \nu}  \,.  
 \label{simpM2} 
\end{align}
where, in the last line, we have used the Maxwell equations, (\ref{Max1}). Note that this last expression precisely matches the term involving $\lambda^I$ in (\ref{Mform2})  but with $\lambda^I \to \alpha^I$.  Indeed, we may rewrite (\ref{Mform2})  as 
\begin{align}
 M ~=~  & -\frac{1}{16 \pi G_5} \,  \int _{ \Sigma}   \nabla_\rho \big[ 2\, Q_{IJ}\,( \lambda^I +\alpha^I) {F^{J \rho \mu}} 
+  Q^{IJ}\, {\Lambda_{I}}_{\sigma} {G_J}^{\,  \sigma \rho \mu }  \big]  \, d \Sigma_{\mu}  \nonumber \\
 ~=~  &  -\frac{1}{16 \pi G_5} \,   \int _{ \Sigma}   \nabla_\rho \big[ 2\, Q_{IJ}\,(Z_I^{-1}) {F^{J \rho \mu}} 
+  Q^{IJ}\, {\Lambda_{I}}_{\sigma} {G_J}^{\,  \sigma \rho \mu }  \big]  \, d \Sigma_{\mu} \,.
\label{Mform3} 
\end{align}
where we have used (\ref{lambdaform}).

We therefore see that the Smarr formula may be arranged, by taking $\alpha^I =1$,  so that the only contribution to the mass is the  topological term  (\ref{Manswer}) and hence (\ref{simpM1}).  On the other hand, one could have chosen $\alpha^I =0$, which kills the explicit topological term but leaves a boundary term at infinity in the original computation. This boundary term is explicitly exhibited in (\ref{simpM2}).  Equation (\ref{Mform3}) gives the exact Smarr formula, independent of the choice of constant in (\ref{lambdaform}).

Taking $\alpha^I =1$ in (\ref{simpM2})  and noting that $Q_{IJ} \to \frac{1}{2} \delta_{IJ}$ at infinity gives:
\begin{equation}
M ~=~    -\frac{1}{16 \pi G_5} \,   \sum_{I=1}^3 \, \int _{\infty}  \,   {F^I}_{\mu \nu} \,   \, d \Sigma^{\mu \nu} ~=~ \frac{1}{8 \pi G_5} \, (2 \pi^2) \, (Q_1 + Q_2 +Q_3)~=~ \frac{\pi}{4  G_5} \, (Q_1 + Q_2 +Q_3) \,,
 \label{BPSres2} 
\end{equation}
where the $2 \pi^2$ comes from the volume of the unit $S^3$.  This result, of course, matches (\ref{MeqlQ}).  
The charges, $Q_I$, are defined by (\ref{ZIexpGH}) and given by (\ref{QIanswer}).  
Given the latter form the charges, it is evident that we could easily obtain the same result from the topological integral.  More specifically, one can rewrite (\ref{simpM1}) as 
\begin{equation}
M ~=~  - \frac{1}{32 \pi G_5}  \, C_{IJK} \,  \alpha^I \,  \int _{ \Sigma}   F^{J}  \wedge F^{K}  \label{simpM3} \,,
\end{equation}
and in Section \ref{FluxesCharges} we will evaluate this by using the intersection form.

\subsection{The topology of the base}

In the light of our discussion in Section \ref{RegTop}, we describe the topology of the base manifold $\cB$ in greater detail.  For more detail, see \cite{Gibbons:1979xm,Yuille,Whitt,Dunajski}.  This is determined by the following data, $\varepsilon_0$, $q_0 = \sum_j q_j $ and $N= \sum_A |q_A|$.  There are essentially three distinct possibilities
\begin{itemize}

\item
If $\varepsilon_0=1$ the base manifold $\{\cB,g_{\mu \nu}\}$  will be  asymptotically  locally flat, ALF, that is near  infinity
the manifold is, in general, a  twisted circle  bundle over $\BR \times S^2$.
If $q_0 = 0$, the bundle is untwisted and the  base manifold $\{B,g_{\mu \nu}\}$  will be asymptotically   flat, AF.

\item
If $\varepsilon_0 =0$ the base manifold $\{\cB,g_{\mu \nu}\}$ 
 will, in general,   be  asymptotically locally  Euclidean, ALE.
That is,  near  infinity the manifold will approach  the quotient 
$\BE^4/C_{|q_0|}$, of Euclidean space   by the  cyclic group of  order $|q_0|$.         
The action of the cyclic group is given by its  embedding into
one of the  $SU(2)$ subgroups  of $SO(4) \equiv ( SU(2) \times SU(2))/\BZ_2$.
If $|q_0|=1$, then base manifold $\{\cB,g_{\mu \nu}\}$  will be asymptotically 
 Euclidean, AE. 

\item
If $\varepsilon_0=0$ and $q_0=0$, then the metric is asymptotic to global $AdS_3 \times S^2$.  
\end{itemize}  

\bigskip
The Euler number, $\chi(\cB) $, is simply the number of $2$-cycles plus $b_0 =+1$ and so:
\begin{equation}
\chi(\cB) = N= \sum_A |q_A| \,.
\end{equation}
A basis $c_A$, $A=1,\dots N-1$ for the second homology, $H_2(\cB,\BZ)$,   given by
an ordering of the points such that the vectors 
$ \vec y^{(A+1)}- \vec y^{(A)}$  and $ \vec y^{(B+1)}- \vec y^{(B)}$ intersect only if 
$|A-B|=1$, and then at only one point. 
The elements of the  intersection form, $I_{AB}$, in this  basis, are 
given by  the cup product 
\begin{equation} 
I_{AB}= c_A \cup c_B \,.
\end{equation}

One has that 
\begin{eqnarray}
I_{AB}&=&\pm |A-B|\,, \quad {\rm if}\qquad |A-B| =1 \nonumber \\
I_{AB}&=& 0 \,, \qquad\qquad  {\rm if} \qquad |A-B| >1\nonumber\\
I_{AA}&=& q_A+ q_{A+1} \,. \label{intmat}
\end{eqnarray}
 The signs of the off-diagonal terms, $I_{AB}$, depend upon the relative orientation of the cycles $A$ and $B$.
  
Then $\tau(\cB)$ is the signature and $\chi(\cB)$ 
the rank plus one of this matrix. Under a change of basis
we have
\begin{equation}
I\rightarrow S^t  I  S \,,\qquad S\in SL(N-1,\BZ)   \,.
\end{equation}

For an axisymmetric configuration, the points lie on a line and this  provides a natural ordering for the cycles.
Thus if $q_j=(+1,+1,-1)$ we have
\begin{equation}
I= \left (\begin{array} {cc} 2 & 1\\  1 &0 \\ \end{array} \right) 
\end{equation}
while if $q_j= (+1,-1+1)$ , we have 
\begin{equation}
I= \left (\begin{array} {cc} 0 & 1\\  1 &0 \\ \end{array} \right) \label{simpint}
\end{equation}
but these are equivalent in $SL(2,\BZ)$ 
since
\begin{equation}
\left (\begin{array} {cc} 1 & -1\\  0 &1 \\ \end{array} \right) 
\left (\begin{array} {cc} 2 & 1\\  1 &0 \\ \end{array} \right) 
\left (\begin{array} {cc} 1 & 0\\  -1 &1 \\ \end{array} \right) 
= \left (\begin{array} {cc} 0 & 1\\  1 &0 \\ \end{array} \right)  \end{equation}
The topology of $\cB$ is thus is $S^2 \times S^2 \setminus \infty$, 
where $\infty$ is the point at infinity which is removed.

More generally, for $N = 2 p +1$ points with $|q_j|=1$ and $q_0=1$, a basis may be found such that the intersection form, $I_{AB}$, is $N$ direct summands of matrices of the form (\ref{simpint}).  Thus the rank is $2p$ and the signature is zero.  Assuming that  manifold to be simply connected, the topology must be the connected sum of $p$ copies of $S^2 \times S^2$ with a point removed.   For $N = 2p$ points with $|q_j|=1$ and $q_0=0$, again assuming simple connectivity, the topology is $(R^2 \times S^2) \# (S^2 \times S^2) \# \dots \# (S^2 \times S^2)$.  Thus every time you add a point and an anti-point you blow up an $S^2 \times S^2$.

\subsection{Fluxes and charges}
\label{FluxesCharges}

As noted earlier, representatives of the homology cycles are given by the $\psi$-fibration over any simple curve between two points,  $\vec y^{(i)}$ and  $\vec y^{(j)}$.  We will thus denote homology cycles by pairs of points.  To introduce a basis we will choose the labeling of the points, $\vec y^{(i)}$, so that 
\begin{equation}
q_j ~=~ (-1)^{j+1} 
\end{equation}
and define a ``simple root'' basis for the homology cycles, $c_A$, by
\begin{equation}
c_A ~=~  \alpha_{A, j}  \vec y^{(j)}\,, \qquad   \alpha_{A,j} ~=~ \delta^j_A  ~-~  \delta^j_{A+1} \,, \qquad A=1, \dots N-1 \,.
\end{equation}
Similarly, the cohomological fluxes can be parametrized by the vectors, $k^I_j$, appearing in the harmonic functions, $K^I$, in (\ref{KVform}).  Indeed, it is convenient to introduce a metric, $\tilde I_{ij}$,  and its inverse, $\tilde I^{ij}$, defined by 
\begin{equation}
\tilde I_{ij}  ~=~   {\rm diag} \,(q_1 , q_2 , \dots, q_N )  \,, \qquad  \tilde I^{ij}  ~=~    {\rm diag}\, (q_1^{-1}, q_2^{-1}, \dots, q_N^{-1}) \,, 
\end{equation}
and then the basis fluxes of (\ref{basicflux}) are given by:
\begin{equation}
\Pi^{(I)} _{A +1\, A}   ~=~  \tilde I^{ij} \,  \alpha_{A, i} \,  k^I_j  \,.
\end{equation}
Define
\begin{equation}
I_{A \, B}   ~=~  \tilde I^{ij} \,  \alpha_{A, i} \,  \alpha_{B, j}   \,,
\end{equation}
and by direct computation one can verify that
\begin{equation}
I_{AB} ~=~ I_{BA} ~=~ 
\begin{cases}
\quad -q_{B}^{-1}  &  \mbox{if} \qquad  B =A+1 \\
\quad -q_{A}^{-1}  &  \mbox{if} \qquad  A =B+1 \\
\quad 0 &  \mbox{otherwise}    \,.\\
\end{cases}
\end{equation}
Comparing with (\ref{intmat}), we see that  $I_{AB}$ is the intersection matrix for our choice of homology basis. Indeed one can make  $I_{A \, A+1} =I_{A+1\, A}=+1$  by reversing the orientations of every second cycle.

Let $\nu^A$ be a dual basis for cohomology, defined by
\begin{equation}
\int _{c_A} \, \nu^B ~=~  \delta^B_A \,, 
\end{equation}
 then we have
\begin{equation}
\int _\cB \, \nu^A  \wedge \nu^B ~=~  I^{AB} ~=~ I^{BA} \label{interint} \,,
\end{equation}
where $I^{AB}$ is the inverse of the intersection matrix, $I^{AB}I_{BC}= \delta ^A_C$.  Suppose that the harmonic parts of the Maxwell fields, $F^I$, are given by 
\begin{equation}
F^I_{harm}  ~=~  \sigma_A^I \, \nu^A \label{harmparts} \,,
\end{equation}
then (\ref{fluxij}) with the normalization set in (\ref{fluxisflux}) implies that 
\begin{equation}
 \sigma_A^I   ~=~4\pi\,  \Pi^{(I)} _{A \, A+1}   ~=~ -4\pi\,    \tilde I^{ij} \,  \alpha_{A, i} \,  k^I_j \,.
\end{equation}
Moreover, (\ref{interint}) implies that the expression (\ref{simpM3}) for the mass may be rewritten
\begin{align}
M ~=~ - \frac{1}{32 \pi G_5}  \, C_{IJK} \,  \alpha^I \,  \int _{ \cB}   F^{J}  \wedge F^{K} & ~=~ - \frac{1}{32 \pi G_5}  \,  C_{IJK} \,  \alpha^I  I^{AB}\,  \sigma_A^J  \,  \sigma_B^K  \nonumber \\
  & ~=~   - \frac{\pi}{2 G_5}  \, C_{IJK} \,  \alpha^I  I^{AB}   (  \tilde I^{ij} \,  \alpha_{A, i} \,  k^J_j)\,  (\tilde I^{k\ell} \,  \alpha_{B,k} \,  k^J_\ell)  \nonumber \\
  & ~=~   - \frac{\pi}{2 G_5} \, C_{IJK} \,\alpha^I\,  \widehat I_{ik}   \,  \tilde I^{k\ell}\,  \tilde I^{ij} k^J_j \,  k^J_\ell  \,,  \label{intFF}
\end{align}
where
\begin{equation}
\widehat I_{ij}~\equiv~  I^{AB} \,  \alpha_{A, i}  \,  \alpha_{B,j} \,.
\end{equation}
However,   observe that 
\begin{align}
\widehat I_{ij} \,  \tilde I^{jk}  \alpha_{A,k}   &~=~ I^{BC} \,  \alpha_{B, i} \,  (\alpha_{C, j}   I^{jk}  \alpha_{A,k}  )  ~=~ I^{BC} \,  \alpha_{B, i} \, I_{CA} ~=~   \alpha_{A,i} \,, \\
\widehat I_{ij} \,  \tilde I^{jk}  q_k &~=~ \sum_{j=1}^N \widehat I_{ij}~=~ I^{AB} \, \alpha_{A, i}  \sum_{j=1}^N   \, \alpha_{B, j} ~=~   0\,.
\end{align}
Thus $\widehat I_{ij} \,  \tilde I^{jk} $ is the projector onto the the space spanned by the $\alpha_{A, i}$ and with null vector $q_k$.  This implies that 
\begin{equation}
\widehat I_{ij} \,  \tilde I^{jk} \,  k^I_k ~=~ \tilde k^I_i   \,,
\end{equation}
where $\tilde k^I_i$ is defined in (\ref{ktilde}). It then follows from  (\ref{intFF}) that
\begin{align}
 M ~=~  &- \frac{1}{32 \pi G_5}  \, C_{IJK} \,  \alpha^I \,  \int _{ \cB}   F^{J}  \wedge F^{K}  \nonumber \\
~=~  & - \frac{\pi}{2 G_5}   \,  C_{IJK} \,\alpha^I\,  \sum_{i=1}^N  \, \frac{ \tilde k^J_i \tilde k^K_i }{q_i}       \nonumber \\
~=~  & \frac{\pi}{4 G_5}   \, \alpha^I\,  Q_I  ~=~ \frac{\pi}{4 G_5}   \,  (Q_1 + Q_2 + Q_3)   \,, 
\end{align}
where we have used (\ref{QIchg}) and taken $\alpha^I = +1$ so that $\lambda^I$ in (\ref{lambdaform}) does indeed fall off at infinity,  which means that there are no boundary terms in the Smarr formula. Once again we reproduce the correct expression, (\ref{MeqlQ}), for the mass and this further confirms our analysis of the topology and the intersection form.

\section{Conclusion}

In this paper we have analyzed the regularity of a class of asymptotically flat BPS fuzzball solutions in five-dimensional supergravity and placed them in the context of the search for gravitational solitons.   In particular, we find that arguments valid in four dimensions based on the Smarr formula, which exclude the existence of solitonic solutions without horizons no longer hold in five dimensions because the Smarr formula acquires extra bulk terms if the spatial manifold has non-trivial second homology. 

Although we constructed solutions using ambi-polar hyper-K\"ahler manifolds, which reverse metric signature on hypersurfaces, the resulting space-times have everywhere non-singular Lorentz metrics.  These singular hypersurfaces become, in the complete Lorentzian metric, a novel form of evanescent ergosurface for which there is no ergo-region.  The space-times are globally hyperbolic and stably causal, having the topology $\IR \times \cB$ where $\cB$ is a complete, four-dimensional Riemannian manifold with non-trivial second homology. Specifically, we find that if $\cB$ is asymptotically Euclidean, {\it i.e.} $q_0=1$, then it has the topology of a $p$-fold connected sum of $(S^2 \times S^2) \# \dots \# (S^2 \times S^2)$ with a point removed and hence has Euler number $2p+1$. If $q_0=0$ then $\cB$ has the topology of $(R^2 \times S^2) \# (S^2 \times S^2) \# \dots \# (S^2 \times S^2)$ with Euler number $2p$, where $p$ is the number of summands.  Thus every time one adds a point and an anti-point one blows up an $S^2 \times S^2$.  It is, perhaps, significant that four-dimensional gravitational instantons representing $p$ black holes, if such instantons exist, are expected to have the foregoing topology.  

It may seem  rather baroque to construct smooth  five-dimensional solutions, by starting from families of four-dimensional singular geometries and then repairing them.  One should however, remember that this approach reduces the BPS equations to a linear system and this means that not only can one construct large families of solitons but one can also analyze their moduli spaces with relative ease.

As our analysis of the Smarr formula shows, the existence of these microstate geometries depends critically upon the non-trivial topology.  The existence of these solitonic geometries also seems to require the Chern-Simons interactions: The solutions are BPS and so necessarily have electric charge and smoothness requires that this must be derived from cohomological fluxes in the spatial topology.  Such fluxes are intrinsically magnetic and so can only generate the electric charge through the Chern-Simons interaction.

There has also been extensive work on extremal non-BPS solutions (see, for example, \cite{Gimon:2007mh,Goldstein:2008fq,Bena:2009ev,Bellucci:2009qv,Bena:2009en,Bena:2009qv,Bena:2009fi,Bobev:2009kn,Dall'Agata:2010dy,Bobev:2011kk,Dall'Agata:2011nh, Vasilakis:2011ki,Bossard:2011kz,Bossard:2012ge})  and our analysis of the Smarr formula is valid for any stationary solution.  It would be interesting to investigate the more detailed analysis to these more general solutions and examine the role of the Chern-Simons interaction for such non-BPS solutions.
  
Finally, there is the question of whether the fuzzball geometries play a role in nature.  Before embarking on such a discussion, one should note that there is considerable danger in trying to extract  properties of general black holes from extremal solutions.  On the other hand, given a BPS soliton, there are  several simple ways to perturb it  to obtain a  near-BPS solitons.   For example, one can include small, non-supersymmetric fluctuations of the bubbles or one can set the elements of the BPS geometry in motion along flat, compact directions of the BPS moduli space.  It is also possible to add stable, perturbative probes that move the mass above the BPS bound    \cite{Bena:2012zi}.  These deformations do not destabilize the original background topology and so the topological support of the soliton is not simply one of extremely fine tuning and it does not evaporate the moment the object is perturbed away from extremality.  It is obviously a huge jump from near-extremality to Schwarzschild black holes, but it is still intriguing to speculate about the role that solitons might play in the universe. 

One of the primary motivations of the fuzzball program is to see if there are geometries that might describe a rich class of semi-classical representatives of black-hole microstates.  Since the majority of the microstates of black holes could potentially involve Planck-scale details that are well beyond the validity of the supergravity approximation, it is most likely that supergravity solitons will only sample the states of the black-hole at some coarse-grained level.  On the other hand, it may just be that such coarse graining will be sufficient to provide a useful, semi-classical description of black-hole thermodynamics.  In this context, it is worth returning, once again, to  historical precedents  and remembering that the Maxwell-Boltzmann distribution and the statistical basis of entropy came 30 years before Planck's constant and over 50 years before the formulation of quantum mechanics.  

More generally, one can interpret that earlier results about ``no solitons without horizons'' as saying that time-independent bosonic fields alone do not seem to be stiff enough to hold up a stationary end state of a star while the more refined result described here shows that there is a new possibility for supporting a stellar remnant: topology and magnetic fluxes.  It is therefore tempting to conjecture new stellar end states, or {\it topological stars}, that are held up by such forces.  The existence of the relevant topological terms in the action requires that space-time have at least five dimensions and that the extra dimensions must therefore begin to play a major role at the horizon scale of an apparently four-dimensional black hole. 

These possibilities may seem far-fetched, but given the unlikely path that led to the discovery of macroscopic fuzzballs in supergravity, one of us at least is very tempted to believe in a seventh impossible thing before breakfast: that fuzz balls will play an important role in nature and in the description of the semi-classical microstate structure of black holes.

\bigskip
\bigskip
\leftline{\bf Acknowledgements}
\smallskip
We would both like to thank KITP for hospitality and a stimulating atmosphere in which this work was initiated. GWG would like to thank the Simons Foundation and the KITP for the award of a Simons Distinguished Visiting Scholarship.   NPW is grateful to the IPhT, CEA-Saclay,  the Institut des Hautes Etudes Scientifiques (IHES), Bures-sur-Yvette and DAMTP, Cambridge for hospitality while this work was completed.  The work of NPW was supported in part by the DOE grant DE-FG03-84ER-40168.  



\end{document}